\documentclass[10pt,twocolumn,amsmath,amssymb,aps,floatfix,preprintnumbers,nofootinbib,nobibnotes,bibnotes]{revtex4-2}
\usepackage{graphicx}
\usepackage{dcolumn}
\usepackage{bm}
\usepackage{hyperref}
\usepackage[mathlines]{lineno}
\usepackage{epsfig}  
\usepackage{color}
\usepackage{slashed}
\usepackage{float}
\usepackage[table]{xcolor}

\def\beq{\begin{equation}}
\def\eeq{\end{equation}}
\def\bea{\begin{eqnarray}}
\def\eea{\end{eqnarray}}

\large

\begin{document}

\title{$CP$ violation in $b \to s \ell \ell$: a model independent analysis}

\author{Neetu Raj Singh Chundawat}
\email{chundawat.1@iitj.ac.in}
\affiliation{Indian Institute of Technology Jodhpur, Jodhpur 342037, India}

\begin{abstract}
We perform a model-independent global fit to all germane and updated $b \to s \ell \ell$ ($\ell=e,\,\mu$) data assuming new physics couplings to be complex. Under the approximation that new physics  universally affects muon and electron sectors and that either one  or two related operators contribute at a time, we identify scenarios which provide a good fit to the data. It turns out that the favored scenarios remain the same as obtained for the real fit. Further, the magnitude of complex couplings can be as large as that of their real counterparts and these  are reflected in the predictions of the direct $CP$ asymmetry, $A_{\rm CP}$,
 in $B \to (K,\, K^*) \mu^+ \mu^-$ along with a number of angular $CP$ asymmetries, $A_i$, in $B^0 \to K^{*0} \mu^+ \mu^-$ decay. The sensitivities of these observables to various solutions are different in the low and high-$q^2$ bins. We also determine observables which can serve as unique identifier for a particular new physics solution. Moreover, we  examine correlations between $A_{\rm CP}$ and several $A_i$ observables. A precise measurement of $A_{\rm CP}$ and $A_i$ observables can not only confirm the existence of additional weak phases but can also enable  unique determination of Lorentz structure of possible new physics in $b \to s \mu^+ \mu^-$ transition.
\end{abstract}
 
\maketitle 

\newpage

\section{Introduction} 
One of the key open problems in particle physics is the observed recalcitrant disparity between the amount of matter and antimatter in the Universe. It is expected that the Big Bang explosion would have created matter and antimatter in equal amounts. However, it is still not understood  how one type of matter triumphed over another in the early Universe. Sakharov's conditions provide  three necessary ingredients required to create the observed baryon asymmetry \cite{Sakharov:1967dj}. One of these conditions requires a $CP$ violation which can be generated through a complex phase in the Lagrangian that cannot be reabsorbed through  the rephasing of the apposite fields.

The SM of electroweak interactions allows for $CP$ violations owing to a complex phase in the quark mixing matrix. The 3 $\times$ 3 CKM matrix can be parametrized by three angles and a single complex phase. This single phase of the CKM matrix is the only source of $CP$ violation in the SM. This phase evinces itself in several observables in the decays of $K$ and $B$ mesons. In fact, the BaBar and Belle experiments established the CKM paradigm of $CP$ violation through several measurements of observables in the decays of $B$ mesons. However, unlike parity violation, which is maximal, the observed $CP$ violation is small and cannot account for the observed baryon asymmetry. The amount of predicted  baryons in the Universe using the CKM formalism falls several orders of magnitude short of the observed value. Therefore one needs to explore  beyond the CKM paradigm of the SM. 

The $CP$ violating observables in the decays induced by the quark level transition $b \to s \mu^+ \mu^-$  are particularly important in probing new
physics. see for e.g. \cite{Kruger:1999xa,Kruger:2000zg,Bobeth:2008ij,Altmannshofer:2008dz,Bobeth:2011gi,Alok:2011gv,Gangal:2022ole,Das:2022xjg,Geng:2022pld,Fleischer:2022klb}. This is because these observables are highly suppressed in the SM. i.e., they are predicted to be less than a percent level in the SM \cite{Kruger:1999xa,Kruger:2000zg}. Even after including the  next-to-leading order 
QCD corrections and hadronic uncertainties, the $CP$ asymmetries are not expected to transcend 1\% \cite{Bobeth:2008ij,Altmannshofer:2008dz,Bobeth:2011gi}.  The $CP$-violating  observables can be measured at LHC or at Belle-II provided new physics enhances them to a level of a few percent. Therefore measurement of any $CP$ violating observable in $b \to s \mu^+ \mu^-$ sector will provide a  luculent signature of  new physics. 

The decay mode $b \to s \mu^+ \mu^-$ is already in spotlight for a decade due to the fact that it has provide a number of observables whose measurements are in contention with the predictions of the SM. These include number of  observables in  $B_s \to \phi\, \mu^+\,\mu^-$ and $B \to K^* \, \mu^+\,\mu^-$  decays which are related only to the muon sector. For e.g., the experimental value of the  branching ratio of $B_s \to \phi\, \mu^+\,\mu^-$ decay  ostentates tension with the SM at 3.5$\sigma$ level \cite{bsphilhc2,bsphilhc3}. The measurement of the optimized angular observable  $P'_5$ in $B \to K^* \, \mu^+\,\mu^-$ decay in the (4.0 $\mathrm{GeV}^2 \le q^2 \le$ 6.0 $\mathrm{GeV}^2$) bin deviates from the SM prediction at the level of 3$\sigma$ \cite{Kstarlhcb1,Kstarlhcb2,LHCb:2020lmf,sm-angular,Matias:2014jua,Hofer:2015kka}{\footnote{ These can also be attributed to under estimation of hadronic uncertainties in the SM such as non factorizable power corrections \cite{Ciuchini:2015qxb,Hurth:2016fbr,Ciuchini:2021smi,Ciuchini:2022wbq}.}}. The measured value of the branching ratio of the decay $B_s \to \mu^+ \mu^-$ also regaled tension with the SM at 2$\sigma$ level \cite{LHCb:2021awg,Combination,ATLAS:2018cur,CMS:2019bbr,LHCb:2017rmj}. However, the CMS collaboration recently updated the measurement of the branching ratio of $B_s \to \mu^+ \mu^-$ \cite{CMS:2022dbz} using the full  Run 2 dataset. This resulted in a new world average of the branching ratio \cite{HFLAV:2022pwe} which is now in agreement with its SM prediction \cite{Bobeth:2013uxa,UTfit:2022hsi}.

 The measurement of the ratio $R_K \equiv \Gamma(B^+ \to K^+ \mu^+ \mu^-)/\Gamma(B^+ \to K^+ e^+ e^-)$  showed a  scantiness of 3.1$\sigma$ as  compared to the SM value in the (1.1 $\mathrm{GeV}^2 \le q^2 \le$ 6.0 $\mathrm{GeV}^2$) bin  \cite{Hiller:2003js,Bordone:2016gaq,LHCb:2021trn}. Here  $q^2$ is the dilepton invariant mass-squared. The   measurements of analogous ratio, $R_{K^*}$,  in the (0.045 $ \rm{GeV}^2 \le q^2 \le$ 1.1 $\rm{GeV}^2$) and (1.1 $\rm{GeV}^2 \le q^2 \le$ 6.0 $\rm{GeV}^2$) bins also nonconcured with the SM at level of $\sim$ 2.5$\sigma$ \cite{rkstar}. In \cite{Isidori:2020acz,Isidori:2022bzw,Nabeebaccus:2022pje}, it was shown that  these deviations are valid even after including  the QED corrections.  These contestations, known as lepton flavour universality violation (LFUV) was accredited to new physics in $b \to s\, \mu^+ \mu^-$ or/and $b \to s\, e^+ e^-$ decays, i.e it required non-universal couplings in muon and electron sectors. However, on  20th of December 2022, the LHCb collaboration updated these measurements \cite{LHCb:2022qnv,LHCb:2022zom} which are now consistent with the SM predictions. As the updated values of $R_K$ and $R_{K^*}$ are now consistent with their SM predictions, these would force the new physics couplings in electron and muon sectors to be {\it nearly} universal in nature.
 
Apart from these LFU ratios, the LHCb collaboration had also provided measurements of  new LFU ratios in $B^0 \to K_S^0 \mu^+  \mu^- $ and $B^+ \to K^{*+} \mu^+ \mu^-$  the channels \cite{LHCb:2021lvy}.   These measurements concur with the SM at $\sim 1.5 \sigma$ level. In the recent updates of  $R_K$ and $R_{K^*}$, the LHCb collaboration included   the experimental systematic effects which were absent in the previous analysis  \cite{LHCb:2022qnv,LHCb:2022zom}. Due to this, the updated values are now consistent with SM predictions. Therefore it is expected that the measurements of  these new LFU observables would also suffer from the same systematic effects. 

 In order to determine the Lorentz structure of possible new physics that can accommodate the anomalous measurements in $b \to s \ell \ell$ decays,  a model independent analysis can be performed using the language of effective field theory \cite{Descotes-Genon:2013wba,Altmannshofer:2013foa,Hurth:2013ssa,Capdevila:2016ivx,Ciuchini:2017mik,Alok:2017jgr,Alok:2019ufo,Altmannshofer:2021qrr,Carvunis:2021jga,Alguero:2021anc,Geng:2021nhg,Hurth:2021nsi,Angelescu:2021lln,Alok:2022pjb,SinghChundawat:2022ldm,Ciuchini:2022wbq}. Barring a few \cite{Alok:2017jgr,Carvunis:2021jga}, most of these analyses assume new physics Wilson Coefficients (WCs) to be real. In this work, we allow the new physics WCs to be complex and perform a global analysis of all  $CP$-conserving $b \to s \ell \ell$ ($\ell=e,\,\mu$) data under the assumption that the beyond SM contributions affect both the muon as well as electron sector equally. Apart from  the updated values of the LFU ratios $R_K$ and $R_{K^*}$ by the LHCb collaboration in December 2022, the branching ratios of  $B \to X_s \mu^+ \mu^-$, $B^0 \to K^0 \mu^+ \mu^-$, $B^+ \to K^+ \mu^+ \mu^-$, $B^0 \to K^{*0} \mu^+ \mu^-$, $B^+ \to K^{*+} \mu^+ \mu^-$ and  $B_s^0 \to \phi \mu^+ \mu^-$ in several $q^2$ bins along  with $B(B_s^0 \to \mu^+ \mu^-)$ are included in the fits. Further, we include a number of $CP$-conserving angular observables in $B^0 \to K^{*0} \mu^+ \mu^-$, $B^+ \to K^{*+} \mu^+ \mu^-$ and $B_s^0 \to \phi \mu^+ \mu^-$ decays.  Moreover, we also include a number of observables in decays induced by $b \to s e^+ e^-$ transition. These observables are obtained by averaging over the angular distributions of $B$ and ${\bar B}$ decays.
 
We take the most frugal approach by considering only one operator or two related operators at a time.  For statistically favoured scenarios, we then obtain predictions for several $CP$-violating observables. For $B^+ \to K^+ \mu^+ \mu^-$ decay, we calculate the direct $CP$ asymmetry, $A_{\rm CP}$, whereas for $B^0 \to K^{*0} \mu^+ \mu^-$  decay,  a number of angular $CP$-asymmetries, $A_i$'s,  are analyzed along with $A_{\rm CP}$. These are obtained by comparing the angular distributions of the corresponding $B$ and ${\bar B}$ decays. For favoured new physics solutions, we also  study correlations between $CP$ violating angular asymmetries in $B^0 \to K^{*0} \mu^+ \mu^-$  decay and  $A_{\rm CP}$  which is expected to be measured with the highest statistical significance amongst all $CP$ asymmetries. These correlations would not only reveal the impact of new physics phase on various quantities   but would also help in sequestering between the allowed scenarios.

 Plan of the work is as follows. In Sec.~\ref{mi}, we discuss the methodology adopted in this work. We then provide the fit results.  Using the fit results, we calculate the direct $CP$ asymmetry in   $B^+ \to K^+ \mu^+ \mu^-$ in Sec.~\ref{pred-acp}.  In the following section, we obtain predictions of a number of $CP$-violating observables in $B^0 \to K^{*0} \mu^+ \mu^-$ decay. We also study correlations between  $A_{CP}$ and several  $CP$-violating angular observables related to $B^0 \to K^{*0} \mu^+ \mu^-$ decay. Finally, we conclude in Sec.~\ref{conc}.

\section{A fit to all $b \to s \ell \ell$ data}
\label{mi}
We start by performing a global fit to all  $CP$ conserving data in $b \to s \ell \ell$ ($\ell=e,\,\mu$) by assuming new physics WCs to be complex.  The data includes the updated measurements of  LFU ratios $R_K$ and $R_K^{*}$ \cite{LHCb:2022qnv,LHCb:2022zom} along with the branching ratio of $B_s \to 
\mu^+ \mu^-$ \cite{HFLAV:2022pwe}.  For reasons mentioned in the Introduction, we do not include measurements of new LFU ratios $R_{K^0_S}$, $R_{K^{* +}}$ \cite{LHCb:2021lvy} in the fit. 
The fit also includes the updated measurements for several $B_s \to \phi \mu^+\mu^-$  observables \cite{bsphilhc3,LHCb:2021xxq}.  
We closely follow the methodology adopted in \cite{Alok:2022pjb} where the new physics couplings were assumed to be real. In this section, we intend to espy the following:
\begin{itemize}
\item The impact of the assumption of the complex coupling on the fit, i.e to spell out the differences between the real and complex fits by making use of the most updated data. 

\item The upper limit on the allowed parameter space of the new weak phases accredited by the current data. This will enable us to identify various $CP$ violating observables where  large enhancement over the SM value is possible. 

\end{itemize}

\subsection{Methodology}

We include  following $CP$ conserving  observables in our fit: 
\begin{enumerate}
\item {\it LFU ratios: } Within the SM, $R_K$ is predicted to be close to unity owing to LFU which is deeply instilled in the symmetry structure of the SM. To be more specific, $R_K^{\rm SM} = 1\pm 0.01$ \cite{Bordone:2016gaq}. This ratio was first measured in 2014 by the LHCb collaboration \cite{LHCb:2014vgu} in   $1.0\, {\rm GeV}^2 \le q^2 \le 6.0 \, {\rm GeV}^2$ bin and updated in 2021  \cite{LHCb:2021trn}. The measured value of $R_K^{\rm exp}=0.846^{+0.044}_{-0.041}$ \cite{LHCb:2021trn} retrogressed from the SM prediction at the level of 3.1$\sigma$. This was considered as an inkling of LFUV.  

In 2017, the notion of  LFUV in $b \to s \ell \ell$  was substantiated by the observation of the ratio $R_{K^*}$ by the LHCb collaboration \cite{rkstar}. This measurement was performed in  two  $q^2$ bins. The measured values
\begin{equation}
R_{K^*}^{\rm exp} =
\left\{
\begin{array}{cc}
0.660^{+0.110}_{-0.070}~ \pm 0.024,&q^2 \subset [0.045,\, 1.1], \\
0.685^{+0.113}_{-0.069}~ \pm 0.047, &q^2 \subset [1.1,\, 6.0], 
\end{array}
\right.
\end{equation}
 detour  
 from the SM predictions \cite{Straub:2018kue,Hiller:2003js} at the level of $\sim$ 2.5$\sigma$. Apart from LHCb, Belle collaboration also measured 
  $R_{K^*}$   in  $0.045 \,{\rm GeV}^{2}< q^2 < 1.1\, {\rm GeV}^2, \, 1.1 \,{\rm GeV}^{2}< q^2 < 6.0\, {\rm GeV}^2, $ and $15.0\, {\rm GeV}^{2}< q^2 < 19.0\, {\rm GeV}^2$ bins \cite{Belle:2019oag}. 

In 2021, LHCb also provided measurements of new LFU ratios $R_{K_S ^0}\equiv \Gamma(B^0 \to K_S^0 \mu^+ \mu^-)/\Gamma(B^0 \to K_S^0 e^+ e^-)$ and $R_{K^{*+}}\equiv \Gamma(B^+ \to K^{*+} \mu^+ \mu^-)/\Gamma(B^+ \to K^{*+} e^+ e^-)$ \cite{LHCb:2021lvy}. The measured value of  $R_{K_S ^0}^{\rm exp} =  0.66^{+0.20+0.02}_{-0.14 - 0.04}$ and 
$R_{K^{*+}}^{\rm exp} =  0.70^{+0.18+0.03}_{-0.13 - 0.04}$ 
in $1.1\, {\rm GeV}^2 \le q^2 \le 6.0 \, {\rm GeV}^2$ bin acquiesces with the SM at 1.5$\sigma$ level \cite{LHCb:2021lvy}. 

On 20th December 2022, the LHCb collaboration provided  updated measurements of the LFU ratios $R_K$ and $R_{K^*}$. The measured values are \cite{LHCb:2022qnv,LHCb:2022zom}:
\begin{equation}
R_{K}^{\rm exp} =
\left\{
\begin{array}{cc}
 0.994^{+0.090}_{-0.082}{\rm (stat)}^{+0.029}_{-0.027}{\rm (sysy)},~~ q^2 \subset [0.1,\, 1.1], \\
0.949^{+0.041}_{-0.041}{\rm (stat)}^{+0.022}_{-0.022}{\rm (sysy)},~~ q^2 \subset [1.1,\, 6.0].
\end{array}
\right.
\end{equation}
\begin{equation}
R_{K^*}^{\rm exp} =
\left\{
\begin{array}{cc}
 0.927^{+0.093}_{-0.087}{\rm (stat)}^{+0.036}_{-0.035}{\rm (sysy)},~~ q^2 \subset [0.1,\, 1.1], \\
1.027^{+0.072}_{-0.068}{\rm (stat)}^{+0.027}_{-0.026}{\rm (sysy)},~~ q^2 \subset [1.1,\, 6.0].
\end{array}
\right.
\end{equation}
It is thus obvious that these values are consistent with their SM predictions. We include these updated measurements in the fit along the Belle measurements of $R_{K^*}$. Further,  $R_{K_S ^0}$ and $R_{K^{*+}}$ measurements are excluded from the fit.

\item {\it Branching ratios:} We include  the updated world average of the  branching ratio of the purely leptonic  decay $ B_s \to \mu^+ \mu^-$ which is $( 3.45 \pm 0.29 ) \times 10 ^{-9} $ \cite{HFLAV:2022pwe}.  This average value is in excellent agreement with the SM prediction \cite{Bobeth:2013uxa,UTfit:2022hsi}. We also consider the branching fractions of inclusive decay modes $B \to X_{s}\mu^{+}\mu^{-}$ and $B \to X_s e^+ e^-$ \cite{Lees:2013nxa}  in the fit in the low and high-$q^2$ bins.

We also ensheathe measurements of the differential branching fraction of several semileptonic decays. The recently updated measurements of the differential branching fraction of $B_s \to \phi \mu^+ \mu^-$ by LHCb
in various  $q^2$ intervals are included in the fit \cite{bsphilhc3}. Further, the differential branching ratios of $B^0 \to K^{*0} \mu^+ \mu^- $ \cite{LHCb:2016ykl,CDFupdate,Khachatryan:2015isa}, $B^{+} \to K^{*+}\mu^{+}\mu^{-}$, $B^{0}\to K^{0} \mu^{+}\mu^{-}$and  $B^{+}\rightarrow K^{+}\mu^{+}\mu^{-}$ \cite{Aaij:2014pli,CDFupdate} in different $q^2$ bins are encapsulated in the analysis. In $b \to s e^+ e^-$ sector, we include  measurement of the differential branching fraction of $B^+ \to K^{+} e^+ e^-$ in $1.0 \le q^2 \le 6.0 ~{\rm GeV}^2$  bin \cite{LHCb:2014vgu}.

\item {\it Angular Observables:} We consider a plentitude  of $CP$ conserving $B^0 \to K^{*0} \mu^+ \mu^- $  angular observables in the fit. This entails longitudinal polarisation fraction $F_L$, forward-backward asymmetry $A_{FB}$  and observables  $S_3$, $S_4$, $S_5$, $S_7$, $S_8$, $S_9$
in various $q^2$ bins, as measured by the LHCb collaboration  \cite{LHCb:2020lmf}. We also include their experimental correlations.  
We also encompass the angular observables $F_L$, $P_1$, $P_4 '$, $P_5 '$, $P_6 '$ and  $P_8 '$ measured by ATLAS \cite{ATLAS:2018gqc} along with $P_1$, $P_5 '$ measured by CMS \cite{CMS:2017rzx}. 
Further, the measurements of $F_L$ and  $A_{FB}$ by CDF and CMS collaborations are also included \cite{ CDFupdate,Khachatryan:2015isa} in our analysis. 

 We then consider full set of angular observables in the decay $B^+ \to K^{*+} \mu^+ \mu^- $ which was measured  for the first time by the LHCb collaboration in 2020 \cite{LHCb:2020gog}.  The  optimized angular observables  $P_1 - P_8 '$ and longitudinal polarisation fraction $F_L$,  along with their experimental correlations are included in the fit \cite{LHCb:2020gog}. Finally, we include $CP$ conserving angular observables in $B_s \to \phi \mu^+ \mu^-$ decay mode. There are $F_L$, $S_3$, $S_4$ and $S_7$ as measured by the LHCb in 2021. The available experimental correlations are also subsumed in the fit \cite{LHCb:2021xxq}. 
 
In decays induced by $b \to s e^+ e^-$ transition, we include the longitudinal polarisation fraction $f_L$ in the decay $B^0 \to K^{*0} e^+ e^-$ in $0.002 \le q^2 \le 1.12 ~{\rm GeV}^2$  bin as measured by the LHCb collaboration \cite{LHCb:2015ycz}. Further, 
we also include  $P'_4$ and $P'_5$ measured by the Belle collaboration in $0.1 \le q^2 \le 4 ~{\rm GeV}^2$, $1.0 \le q^2 \le 6.0 ~{\rm GeV}^2$ and $14.18 \le q^2 \le 19.0 ~{\rm GeV}^2$ bins \cite{Belle:2016fev}.

\end{enumerate}

\begin{table*}[htb]
  \begin{center}
\begin{tabular}{|c||c|c||c|c|}
\hline\hline
Wilson Coefficient(s) & \multicolumn{2}{|c|}{{\tt Real}} & \multicolumn{2}{|c|}{Complex} \\ \hline
  & Best fit value(s) & $\Delta \chi^2_{\rm real}$ & $1\sigma$ range [Re($C_i$), Im($C_i$)] & $\Delta \chi^2_{\rm complex}$ \\  
\hline
$C_i=0\,\,\rm (SM)$ & - & 0 & - & 0 \\ 
\hline \hline
1D Scenarios: & & & &\\\hline		
$C_9^{\rm NP}$ & $-1.08 \pm 0.18$  & 27.90 & [(-1.34, -0.80), (-0.86, 0.93)]   &27.91\\ 
\hline 
$C_{10}^{\rm NP}$   & $0.35 \pm 0.15 $  & 5.80 & [(0.24, 0.99), (-2.09, 2.08)]&   7.64\\ 
\hline  
$C_9^{\rm NP} = -C_{10}^{\rm NP}$   & $-0.50 \pm 0.12 $  &  18.85 & [(-0.83, -0.32), (-1.21, 1.31)] &  18.91 \\
\hline 
$C_9^{\rm NP} = - C_{9}^{'}$    & $0.88 \pm 0.16$  & 26.92 & [(-1.12, -0.66), (-0.88, 0.89)] &   28.10 \\ 
\hline \hline
\end{tabular}
\caption{The best fit values of new WCs in various 1D scenarios. Here $\Delta\chi^2 = \chi^2_{\rm SM}-\chi^2_{\rm bf}$ where $\chi^2_{\rm bf}$ is the $\chi^2$ at the best fit point and $\chi^2_{\rm SM}$  corresponds to the SM which is $\chi^2_{\rm SM}  \approx$ 184.}
\label{fit-1}
 \end{center}
\end{table*}

\begin{figure*}[htb]
\centering
\includegraphics[width = 3.1in]{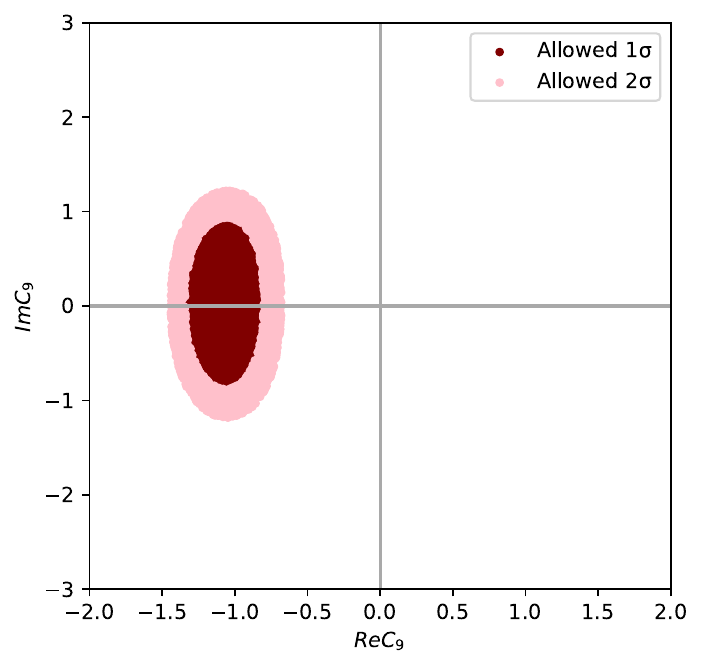}
\includegraphics[width = 3.1in]{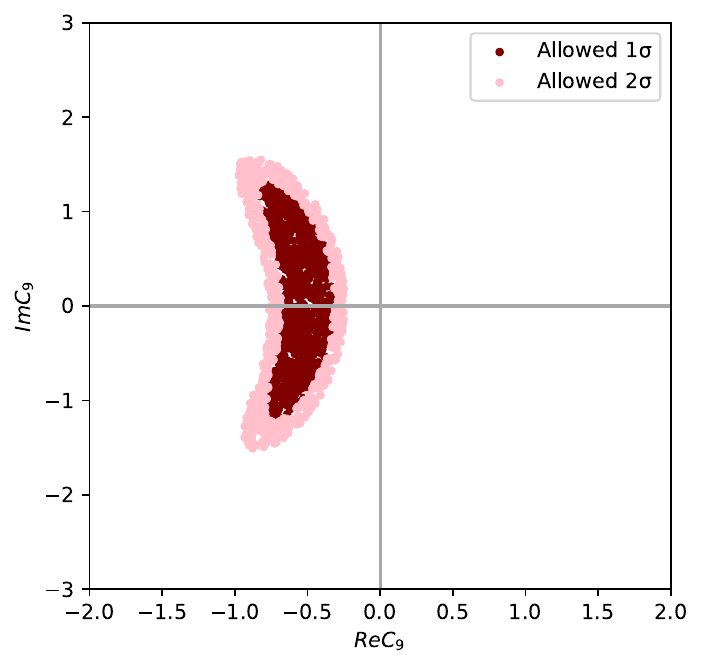}\\
\includegraphics[width = 3.1in]{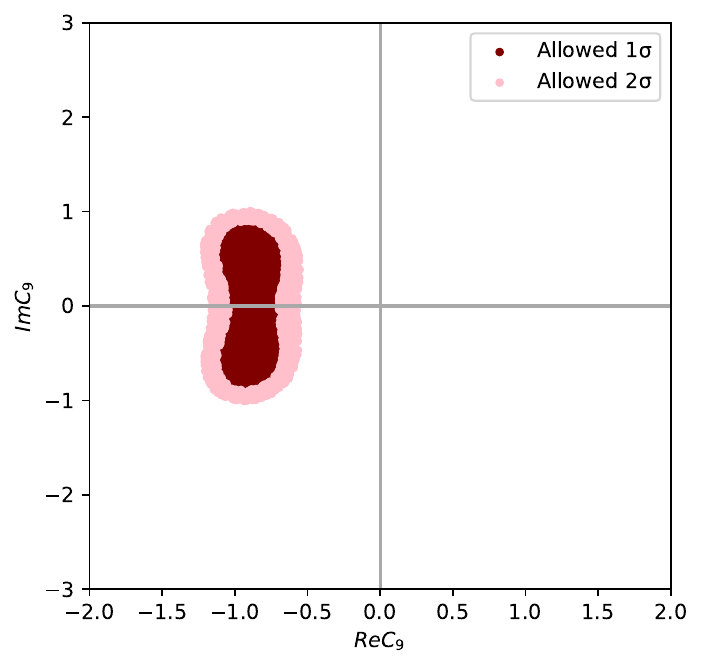}
\caption{Allowed parameter space for new physics Scenarios  $C_9^{\rm NP}$ (upper left panel), $C_9^{\rm NP}=-C_{10}$ (upper right panel) and $C_9^{\rm NP}=-C_{9}^{'}$ (lower panel). }
\label{fig:contour}
\end{figure*}

In order to identify the Lorentz structure of possible new physics that can account for the discrepancies in $ b \to s \ell \ell$ data, we perform a model independent analysis within the framework of effective field theory. For this we consider new physics in the form of vector and axial-vector operators.  The effective Hamiltonian for $ b \to s \ell \ell$ transition is then given by 
\begin{eqnarray}
  \mathcal{H}_{\mathrm{eff}}(b \rightarrow s \ell \ell) =
  \mathcal{H}^{\rm SM} + \mathcal{H}^{\rm VA}  \; .
\label{heff}
\end{eqnarray}
Here the SM effective Hamiltonian can be written as 
\begin{align} \nonumber
  \mathcal{H}^{\rm SM} &= -\frac{ 4 G_F}{\sqrt{2} \pi} V_{ts}^* V_{tb}
  \bigg[ \sum_{i=1}^{6}C_i \mathcal{O}_i + C_8 {\mathcal O}_8 \nonumber\\
    & + C_7\frac{e}{16 \pi^2}[\overline{s} \sigma_{\alpha \beta}
      (m_s P_L + m_b P_R)b] F^{\alpha \beta}  \nonumber\\
    & + C_9^{\rm SM} \frac{\alpha_{\rm em}}{4 \pi}  
    (\overline{s} \gamma^{\alpha} P_L b)(\overline{\ell} \gamma_{\alpha} \ell)  
  \nonumber\\
    &  + C_{10}^{\rm SM} \frac{\alpha_{\rm em}}{4 \pi}
    (\overline{s} \gamma^{\alpha} P_L b)(\overline{\ell} \gamma_{\alpha} \gamma_{5} \ell)
    \bigg] \; ,
\end{align}
where  $V_{ij}$ are the elements of the Cabibbo-Kobayashi-Maskawa
(CKM) matrix. The short-distance contributions are enciphered in the WCs $C_i$ of the four-fermi operators ${\cal O}_i$ where the scale-dependence  is implicit, i.e. $C_i \equiv C_i(\mu)$
and ${\cal O}_i \equiv {\cal O}_i(\mu)$. 
The operators ${\cal O}_i$ ($i=1,...,6,8$) contribute
through the modifications  $C_7(\mu) \rightarrow C_7^{\mathrm{eff}}(\mu,q^2)$
and $C_9(\mu) \rightarrow C_9^{\mathrm{eff}}(\mu,q^2)$.
The new physics effective Hamiltonian can be written as
\begin{align} \nonumber
  \mathcal{H}^{\rm VA} &=
  -\frac{\alpha_{\rm em} G_F}{\sqrt{2} \pi} V_{ts}^* V_{tb}
  \bigg[C_9^{\rm NP} (\overline{s} \gamma^{\alpha} P_L b)
    (\overline{\ell} \gamma_{\alpha} \ell) \nonumber\\
    &
    + C_{10 }^{\rm NP} (\overline{s} \gamma^{\alpha} P_L b)
    (\overline{\ell} \gamma_{\alpha} \gamma_{5} \ell)  \nonumber\\ &
    + C_9^{\prime}(\overline{s} \gamma^{\alpha} P_R b)(\overline{\ell} \gamma_{\alpha} \ell)  \nonumber\\
    &
    + C_{10 }^{\prime} (\overline{s} \gamma^{\alpha} P_R b)
    (\overline{\ell} \gamma_{\alpha} \gamma_{5} \ell) \bigg]\,.
    \label{heff-va}
\end{align}
Here $C_9^{\rm NP},\, C_{10}^{\rm NP},\, C_9^{\prime}$ and $C_{10}^{\prime}$ are the new physics WCs which are assumed to be complex in the current analysis. Following a penurious approach, we  ruminate only those scenarios where either only one new physics operator  or two operators whose WCs are linearly related, contributes. We call them as ``1D" scenarios. Under this assumption, we perform a  $\chi^2$ fit to  identify solutions which can accommodate the current $ b \to s \ell \ell$ measurements. The fit is performed using the CERN minimization code {\tt MINUIT} \cite{James:1975dr}. The  $\chi^2$ which is a function of new physics WCs is defined as 
\begin{equation}
  \chi^2(C_i,C_j) = \big[\mathcal{O}_{\rm th}(C_i,C_j) -\mathcal{O}_{\rm exp}\big]^T \,
  \mathcal{C}^{-1} \, \big[\mathcal{O}_{\rm th}(C_i,C_j) -\mathcal{O}_{\rm exp} \big]\,,
  \label{chisq-bsmumu}
\end{equation} 
where $\mathcal{O}_{\rm th}(C_i,C_j)$ are the theoretical predictions of the N=179 observables used in the fit and $\mathcal{O}_{\rm exp}$ are the corresponding central values of the experimental measurements. The total $N \times N$  covariance matrix is obtained by adding the individual  theoretical and experimental covariance matrices.
The theoretical predictions of N=179 observables along with the theoretical covariance matrix are evaluated using {\tt flavio} \cite{Straub:2018kue} {where the observables are preimplemented based on  refs. \cite{Bharucha:2015bzk,Gubernari:2018wyi}.    The experimental  correlations, $\mathcal{O}_{\rm exp}$, are admitted for  the angular observables in $B^0 \to K^{*0} \mu^+ \mu^-$ \cite{LHCb:2020lmf}, $B^+ \to K^{*+} \mu^+ \mu^-$ \cite{LHCb:2020gog} and $B_s \to \phi \mu^+ \mu^-$ \cite{LHCb:2021xxq}. Further, for asymmetric errors,  we use the larger error on both sides of the central value. 

The $\chi^2$  value in the SM is represented by $\chi^2_{\rm SM}$ whereas $\chi^2_{\rm bf}$ represents the value at the  the best-fit point in the presence of new physics. We then quantitate the goodness of fit by  $\Delta\chi^2 \equiv \chi^2_{\rm SM}-\chi^2_{\rm bf}$ for each new physics scenario. This means that, under the given assumptions, the largest value of this quantity would represent the best possible new physics scenario to accommodate the entire $b \to s \ell \ell$ data.

\subsection{Fit results}
The fit results are presented in Table \ref{fit-1}. For comparison, we  provide the updated fit results for the real WCs. Using the values of $R_K$ and $R_{K^*}$ along with the measurement of LFU ratios $R_{K^0_S}$ \& $R_{K^{* +}}$ \cite{LHCb:2021lvy}  and older world average of the branching ratio of $B_s \to \mu^+ \mu^-$ \cite{Altmannshofer:2021qrr}, it was well established that for real WCs, the new physics solutions $C_9^{\rm NP}$ and  $C_9^{\rm NP}=-C_{10}$ provided a good fit to the data whereas $C_9^{\rm NP}=-C_{9}^{'}$ scenario provided a moderate fit, see for e.g. \cite{Alguero:2021anc,Alok:2022pjb}. In \cite{Alok:2022pjb}, it was shown that $C_{10}^{\rm NP}$ scenario also provided a moderate fit to the data at par with $C_9^{\rm NP}=-C_{9}^{'}$ solution. It is perspicuous from Table \ref{fit-1} that the updated fit for real WCs still prefers $C_9^{\rm NP}$ scenario. The $C_9^{\rm NP}=-C_{9}^{'}$ solution which provided a moderate fit to the older data now provides a good fit at par with  $C_9^{\rm NP}$ solution. However, the value of 
$\Delta\chi^2$ for  $C_9^{\rm NP}=-C_{10}$  scenario falls considerably, $\sim 10$ below $\Delta\chi^2$ for $C_9^{\rm NP}$ solution. Therefore the   $C_9^{\rm NP}=-C_{10}$  scenario can only provide a moderate fit to the current $b \to s \ell \ell$ data. The situation appears to be more grim for $C_{10}^{\rm NP}$ scenario  which fails to provide useful improvement in the value of $\chi^2$ as compared to the  $\chi^2_{\rm SM}$. This is mainly due to the fact the current world average of the branching ratio of $B_s \to \mu^+ \mu^-$ is now in excellent agreement with the SM value.

It is also apparent  from Table \ref{fit-1} that the scenarios that are favoured by assuming new physics WCs to be real, remains the preferred ones even for the complex couplings. The $C_9^{\rm NP}$ and $C_9^{\rm NP}=-C_{9}^{'}$ scenarios turn out to be the most viable scenarios to accommodate all $b \to s \ell \ell$ data whereas the $C_9^{\rm NP}=-C_{10}$ scenario can only provide a moderate fit to the current data. The $C_{10}^{\rm NP}$ scenario has the lowest value of $\Delta \chi^2$ in comparison to the other three solutions. Hence we drop this from further consideration in this work. Further,  the imaginary part of all WCs  are allowed to have  values similar to that of their real counterparts. This is evident from the 1$\sigma$ range of complex WCs shown in Fig.~\ref{fig:contour}. Therefore, it will be intriguing to see whether some of the $CP$ violating observables can be enhanced up to the current or planned sensitivity of LHCb or Belle-II.

\section{Direct $CP$ asymmetry in $B^+ \to K^+ \mu^+ \mu^-$}
\label{pred-acp}

The $CP$ violation can be classified into two types: the direct $CP$ violating asymmetries and triple product $CP$ asymmetries. Assume  that there are two contributions, $A_1 \propto e^{i\alpha_1} e^{i\beta_1}$ and $A_2 \propto e^{i\alpha_2} e^{i\beta_2}$, in $b \to s \mu^+ \mu^-$ decay. Here $\alpha_{1,2}$ and $\beta_{1,2}$ are weak and strong phases, respectively. It can be easily shown that the direct $CP$ asymmetries are proportional to $\sin(\alpha_1- \alpha_2)\, \sin(\beta_1- \beta_2)$. This implies that these asymmetries can have non-zero values only if the two interfering amplitudes have a relative  weak as well as strong phase. 
  On the other hand, as triple product asymmetries are proportional to $\sin(\alpha_1- \alpha_2)\, \cos(\beta_1- \beta_2)$,   a relative weak phase between the amplitudes is sufficient to provide a non-zero value. 
  
  \begin{table}[ht]
  \begin{center}
\begin{tabular}{|c||c|c|}
\hline\hline  Wilson Coefficients   &  $A^K_{{\rm CP} {[1-6]}}\, (\%)$   & $A^K_{{\rm CP} {[15-19]}}\, (\%)$   \\ \hline
$C_i=0\,\,\rm (SM)$ & $\approx$ 0 & $\approx$ 0    \\ 
\hline \hline
1D Scenarios:  &  &    \\ \hline		
$C_9^{NP}$ & (-0.33, 0.58) & (-3.53, 3.54) \\ 
\hline  
$C_9^{NP} = -C_{10}^{NP}$ &(-0.52, 0.85) & (-5.24, 5.35)\\ 
\hline 
$C_9^{NP} = - C_{9}^{'}$   &( 0.12, 0.12) & (-0.16, -0.16)\\ 
\hline \hline
\end{tabular}
\caption{Predictions  of  $A_{\rm CP}$ in $B^+ \to K^+ \mu^+ \mu^-$  decay (1$\sigma$ range). Here $A^K_{{\rm CP}} \equiv A_{\rm CP} (B^+ \to K^+ \mu^+ \mu^-)$. }
\label{pred-acp0}
 \end{center}
\end{table}

  For  $B \to K  \mu^+ \mu^-$  decays, we only have direct $CP$ asymmetry, $A_{CP}$,  which is defined as 
  \begin{equation}
  A_{CP} = \frac{\Gamma-\bar{\Gamma}}{\Gamma+\bar{\Gamma}}\,
  \end{equation}
  where $\Gamma$ and $\bar{\Gamma}$ are the decay rates  of $B^+ \to K^+ \mu^+ \mu^-$ and $\bar{B} \to \bar{K} \mu^+ \mu^-$ decays. $\bar{\Gamma}$ is obtained from  $\Gamma$  by changing the sign of the weak phases. The sign of strong phases remain unchanged. The theoretical expression of $\Gamma(B \to K \mu^+ \mu^-)$ is provided in Appendix \ref{appen0}.

Effectively, various contributions to $b \to s \mu^+ \mu^-$ decay are proportional to CKM factors $V^*_{tb}V_{ts}$ and $V^*_{ub}V_{us}$. The term proportional to $V^*_{cb}V_{cs}$ is eliminated using the unitarity of the CKM mixing matrix. Although the phase of $V^*_{ub}V_{us}$ is large but its magnitude is suppressed in comparison to $V^*_{tb}V_{ts}$. Therefore, within the SM, various contributions to $b \to s \mu^+ \mu^-$ have almost similar weak phase. Further, the $c\bar{c}$
and $u\bar{u}$ quark loop generate strong phase in the WC $C_9^{\rm eff}$. However, this is not substantial and hence the $CP$ asymmetries in $b \to s \mu^+ \mu^-$ are highly suppressed within the SM. The fact that the new physics strong phase is negligibly small, one requires a large value of new physics weak phase to provide an enhancement at the level of few percent \cite{dattlon}. 

Including the dielectron mode, the Belle and BaBar collaborations measured direct $CP$ asymmetries in  $B^+ \to K^+ \ell \ell$ and $B^0 \to K^{*0} \ell \ell$ decay modes. Based on a data sample of 657 million $B\bar{B}$ pairs, Belle reported $A_{\rm CP}$ in $B^+ \to K^+ \ell \ell$ and $B^0 \to K^{*0} \ell \ell$ to be $0.04 \pm 0.10$ and $-0.10 \pm 0.10$,  respectively \cite{Belle:2009zue}. These values for BaBar collaboration are $-0.03 \pm 0.14$ and $0.03 \pm 0.13$, respectively \cite
{BaBar:2012mrf}. This corresponds to a sample of 471 million $B\bar{B}$ events. These asymmetries, in the muonic channel, were measured by the LHCb collaboration by making use of a data set corresponding to an integrated luminosity of 3.0 $\rm fb^{-1}$ collected in 2011 and 2012 at centre of mass energies of 7 and 8 TeV , respectively. The measured values are $A_{\rm CP}(B^+ \to K^+ \ell \ell)=0.012 \pm 0.017$ and  $A_{\rm CP}(B^0 \to K^{*0} \ell \ell)=-0.035 \pm 0.024$ 
\cite{LHCb:2014mit,LHCb:2012kz,LHCb:2013lvw}. The quoted values are in the full-$q^2$ region. Owing to large errors, these measurements are consistent with the SM. But on the other hand, a possibility of $A_{\rm CP}$ at a level of a few percent is not ruled out. It would be interesting to see whether such an enhancement is allowed by the current data. 

The predictions of  $A_{\rm CP}$ in $B^+ \to K^+ \mu^+ \mu^-$ for all 1D favored solutions are given in Table~\ref{pred-acp0}. It is apparent that none of the new physics solutions can enhance $A^K_{{\rm CP}}$ in the low-$q^2$ bin at the level of a few percent. However, such an enhancement is feasible in the high-$q^2$ region for $C_9^{NP}$ and $C_9^{NP} = -C_{10}^{NP}$ solutions, the enhancement being more prominent for the later solution. Therefore any observation of $A_{\rm CP}$ in $B^+ \to K^+ \mu^+ \mu^-$ can be attributed to either of these solutions. The $C_9^{NP} = - C_{9}^{'}$  scenario predicts $A^K_{{\rm CP}} < 1\%$.  

 Here one should emphasize that although the enhancement in the high-$q^2$ bin is more prominent, the measurement of $A_{\rm CP}$ in the low-$q^2$ region appears to be more attractive as the branching ratio in the low-$q^2$ region is larger as compared to the high-$q^2$ bin.  Belle-II experiment is expected to collect a sample of a few thousand events of $B \to (K,\,K^*) \mu^+ \mu^-$ \cite{Belle-II:2018jsg}. With such an event sample, it would be  possible to have a 3$\sigma$ determination of the $CP$ asymmetries
which are of a few percent level.

\section{$CP$ violating observables in $B^0 \to K^{*0} \mu^+ \mu^-$}
\label{pred-kstar}

The differential distribution of $B\to K^*(\to K\pi)\mu^+\mu^-$ decay can be parametrized in terms of  one kinematic and three angular variables. The kinematic variable is $q^2 = (p_B-p_{K^*})^2$, where $p_B$ and $p_{K^*}$ are the four-momenta of $B$ and $K^*$ mesons, respectively. The angular variables are usually  defined in the rest frame of the vector meson $K^*$. These angles are  \begin{itemize}
\item  $\theta_{K}$ the angle between 
$B$ and $K$ mesons where $K$ meson emerges from the decay of a $K^*$,

\item $\theta_{\mu}$ the angle between $\mu^-$ and $B$ momenta,

\item  $\phi$ the angle between $K^*$ decay plane and the plane defined by the $\mu^+-\mu^-$ momenta.
\end{itemize}

 \begin{table}[htb]
  \begin{center}
\begin{tabular}{|c||c|c|}
\hline\hline  Wilson Coefficients     & $A^{K^*}_{{\rm CP} {[1-6]}}\, (\%)$ & $A^{K^*}_{{\rm CP} {[15-19]}}\, (\%)$ \\ \hline
$C_i=0\,\,\rm (SM)$ & $\approx$ 0 & $\approx$ 0   \\ 
\hline \hline
1D Scenarios:  &  &    \\ \hline		
$C_9^{NP}$  & (0.01, 0.11) & (-1.79, 1.78)\\ 
\hline  
$C_9^{NP} = -C_{10}^{NP}$  & (-0.04, 0.14) & (-2.69, 2.71)\\ 
\hline 
$C_9^{NP} = - C_{9}^{'}$   & (-0.06, 0.14) & (-3.19, 3.09)\\ 
\hline \hline
\end{tabular}
\caption{Predictions  of  $A_{\rm CP}$ in  $B^0 \to K^{*0} \mu^+ \mu^-$ decay (1$\sigma$ range). Here  $A^{K^*}_{\rm CP} \equiv A_{\rm CP} (B^0 \to K^{*0} \mu^+ \mu^-)$. }
\label{pred-acp1}
 \end{center}
\end{table}

\begin{table*}[htb]
  \begin{center}
  \resizebox{\textwidth}{!}{ 
\begin{tabular}{|c||c|c|c|c|c|c|c|}
\hline\hline
Wilson Coefficient(s)  &  ${A_3}_{[1-6]} \, (\%)$  &  ${A_4}_{[1-6]} \, (\%)$ &   ${A_5}_{[1-6]}\, (\%)$ & ${A_6^s}_{[1-6]}\, (\%)$ &    ${A_7}_{[1-6]}\, (\%)$ & ${A_8}_{[1-6]}\, (\%)$ &${A_9}_{[1-6]}\, (\%)$
\\ \hline
$C_i=0\,\,\rm (SM)$ & $\approx$ 0  &  $\approx$ 0 &  $\approx$ 0& $\approx$ 0 & $\approx$ 0 & $\approx$ 0  &  $\approx$ 0  \\ 
\hline \hline
1D Scenarios: &   & & && &&\\ \hline		
$C_9^{NP}$ & (0.00, 0.01)  & (-0.10, 0.15) & (0.04, 0.04) & (-0.08, -0.07)  & (0.27, 0.29) & (-3.03, 3.45)  & (-0.31, 0.36)\\ 
\hline 
$C_9^{NP} = -C_{10}^{NP}$  & (-0.01, 0.02)  & (-0.17, 0.23) & (-0.40, 0.49) & (-0.79, 0.60) & (-9.17, 10.38) &  (-4.74, 5.34) & (-0.49, 0.55)\\
\hline 
$C_9^{NP} = - C_{9}^{'}$   & (-0.24, 0.23)  &   (-0.24, 0.28)&  (0.04, 0,04)& (-0.09, -0.08) & (0.27, 0.32) & (-0.91, 1.04 ) &  (-0.21, 0.15)\\ 
\hline \hline
\end{tabular}
}
\caption{Prediction  of various $CP$ violating angular observables (1$\sigma$ range) in $B^0 \to K^{*0} \mu^+ \mu^-$ in the low-$q^2$ region. }
\label{pred-1}
 \end{center}
\end{table*}

   \begin{table*}[ht]
  \begin{center}
  \resizebox{\textwidth}{!}{ 
\begin{tabular}{|c||c|c|c|c|c|c|c|}
\hline\hline
Wilson Coefficient(s)  &  ${A_3}_{[15-19]} \, (\%)$  &  ${A_4}_{[15-19]} \, (\%)$ &   ${A_5}_{[15-19]} \, (\%)$ & ${A_6^s}_{[15-19]} \, (\%)$ &    ${A_7}_{[15-19]} \, (\%)$& ${A_8}_{[15-19]}\, (\%)$ &${A_9}_{[15-19]}\, (\%)$
\\ \hline
$C_i=0\,\,\rm (SM)$ & $\approx$ 0  &  $\approx$ 0 &  $\approx$ 0& $\approx$ 0 & $\approx$ 0 & $\approx$ 0  &  $\approx$ 0  \\ 
\hline \hline
1D Scenarios: &  &   & && &&\\ \hline		
$C_9^{NP}$  &  (-0.72, 0.73) & (-1.06, 1.07)  & (0.10, 0.11)  & (-0.21, -0.18)  & (0.01, 0.01) & (-0.13, 0.15)  & (-0.11, 0.13)\\ 
\hline 
$C_9^{NP} = -C_{10}^{NP}$  & (-1.11, 1.10)  & (-1.62, 1.61) & (-1.51, 1.58) & (-2.79, 2.65)  & (-0.40, 0.45) &  (-0.20, 0.24) & (-0.17, 0.20)\\
\hline 
$C_9^{NP} = - C_{9}^{'}$  & (-1.91, 1.91)  & (-2.13, 2.17) & (0.10, 0.12) & (-0.22, -0.18) & (0.01, 0.01)& (-3.58, 3.56) & (-6.40, 6.33) \\ 
\hline \hline
\end{tabular}
}
\caption{Prediction  of various $CP$ violating angular observables (1$\sigma$ range) in $B^0 \to K^{*0} \mu^+ \mu^-$ in the high-$q^2$ region. }
\label{pred-2}
 \end{center}
\end{table*}

 The four-fold decay distribution can be expounded as~\cite{Bobeth:2008ij,Altmannshofer:2008dz}
\begin{equation}
\frac{d^4\Gamma}{dq^2d\cos\theta_{\mu}d\cos\theta_{K}d\phi} = \frac{9}{32\pi}I(q^2,\theta_{\mu},\theta_{K},\phi),
\end{equation}
where
\begin{eqnarray}
I(q^2,\theta_{\mu},\theta_{K},\phi) &=& I^s_1\sin^2\theta_{K} + I^c_1\cos^2\theta_{K}\nonumber\\
& &+(I^s_2\sin^2\theta_{K} 
+I^c_2\cos^2\theta_{K})\cos 2\theta_{\mu} \nonumber\\
& & +I_3\sin^2\theta_{K}\sin^2\theta_{\mu}\cos 2\phi \nonumber\\
& & +I_4\sin 2\theta_{K}\sin 2\theta_{\mu}\cos\phi \nonumber\\
& & + I_5 \sin 2\theta_{K}\sin\theta_{\mu}\cos\phi \nonumber \\
& & + I^s_6\sin^2\theta_{K} \cos\theta_{\mu}\nonumber\\
& & + I_7\sin 2\theta_{K}\sin\theta_{\mu}\sin\phi \nonumber\\
& & + I_8\sin 2\theta_{K}\sin 2 \theta_{\mu} \sin\phi \nonumber\\
& &+I_9\sin^2\theta_{K}\sin^2\theta_{\mu}\sin 2\phi\,.
\label{Ifunc}
\end{eqnarray}

The  expressions of these twelve angular coefficients $I^{(a)}_i$ \cite{Kruger:1999xa,Altmannshofer:2008dz,Gratrex:2015hna} are provided  in Appendix \ref{appen}. 
These  coefficients  depend on the $q^2$ variable and on various hadronic form factors.
 The corresponding expression for the four-fold decay distribution  of the  $CP$ conjugate decay mode can be obtained by substituting $\theta_{\mu}$ by $(\pi-\theta_{\mu})$ and $\phi$ by $-\phi$. This results in the following transformations of angular coefficients
\begin{equation}
I^{(a)}_{1,2,3,4,7} \Longrightarrow \bar{I}^{(a)}_{1,2,3,4,7}, \quad I^{(a)}_{5,6,8,9} \Longrightarrow -\bar{I}^{(a)}_{5,6,8,9}\,.
\end{equation}
Here  $\bar{I}^{(a)}_i$ are the complex conjugate of $I^{(a)}_i$. Therefore, one can define twelve $CP$ averaged angular observables  as~\cite{Bobeth:2008ij,Altmannshofer:2008dz}
\begin{equation}
S^{(a)}_i(q^2) = \frac{I^{(a)}_i(q^2)+ \bar{I}^{(a)}_i(q^2)}{d(\Gamma +\bar{\Gamma})/dq^2},
\end{equation}
along with twelve $CP$ asymmetries 
\begin{equation}
A^{(a)}_i(q^2) = \frac{I^{(a)}_i(q^2)- \bar{I}^{(a)}_i(q^2)}{d(\Gamma +\bar{\Gamma})/dq^2}\,.
\end{equation}
The $CP$ asymmetry in the dimuon mass spectrum is defined as
\begin{equation}
A_{\rm CP} (q^2) = \frac{d\Gamma/dq^2 - d\bar{\Gamma}/dq^2}{d\Gamma/dq^2 + d\bar{\Gamma}/dq^2},
\end{equation}
where $d\Gamma/dq^2$ can be expressed in terms of angular coefficients as
\begin{equation}
\frac{d\Gamma}{dq^2} = \frac{3}{4}\left(2I_1^s + I_1^c\right) - \frac{1}{4}\left(2I_2^s + I_2^c\right)\,.
\end{equation}

\newcommand{\img}{\includegraphics[scale=0.10]{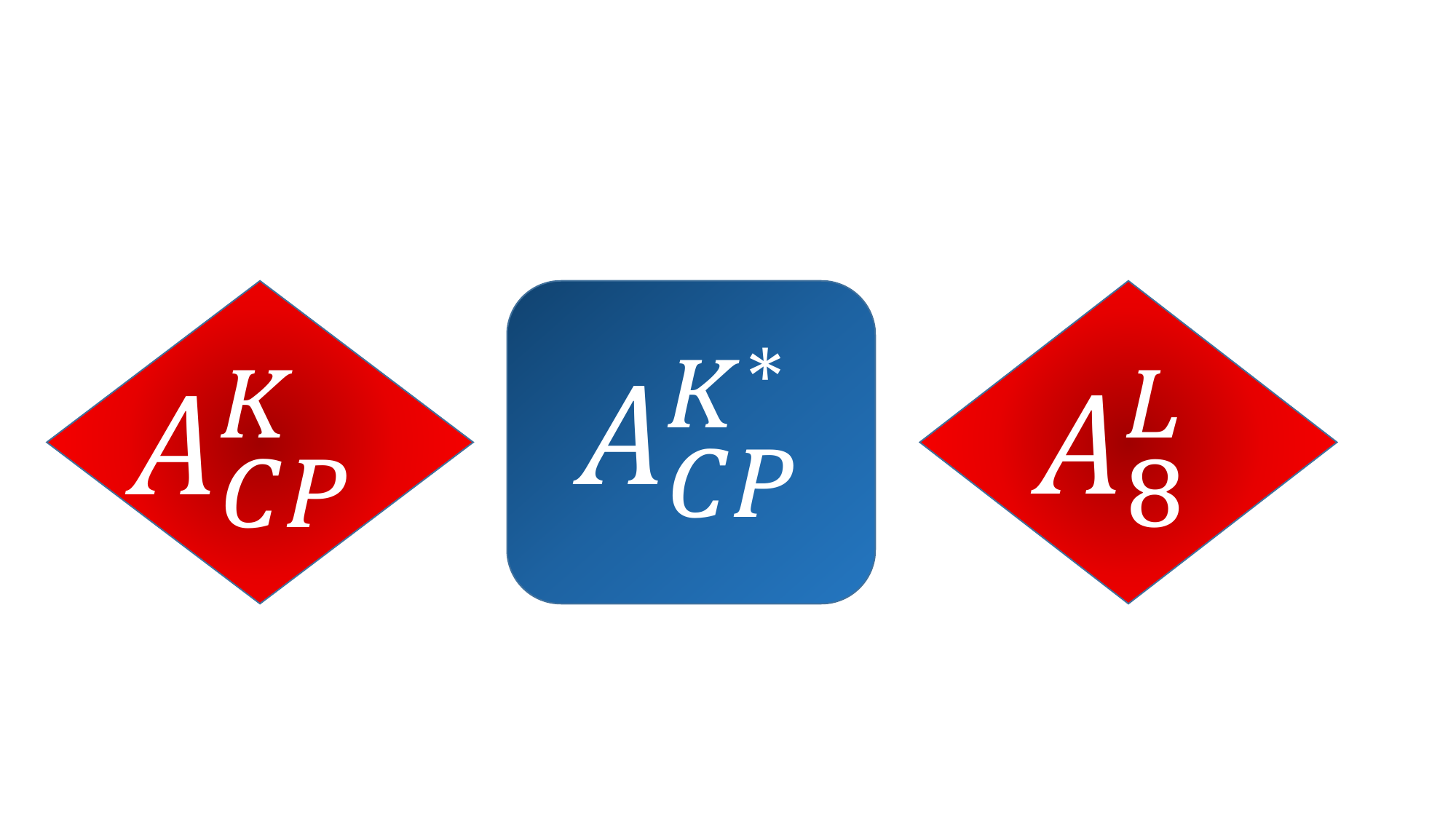}}
\newcommand{\imgone}{\includegraphics[scale=0.10]{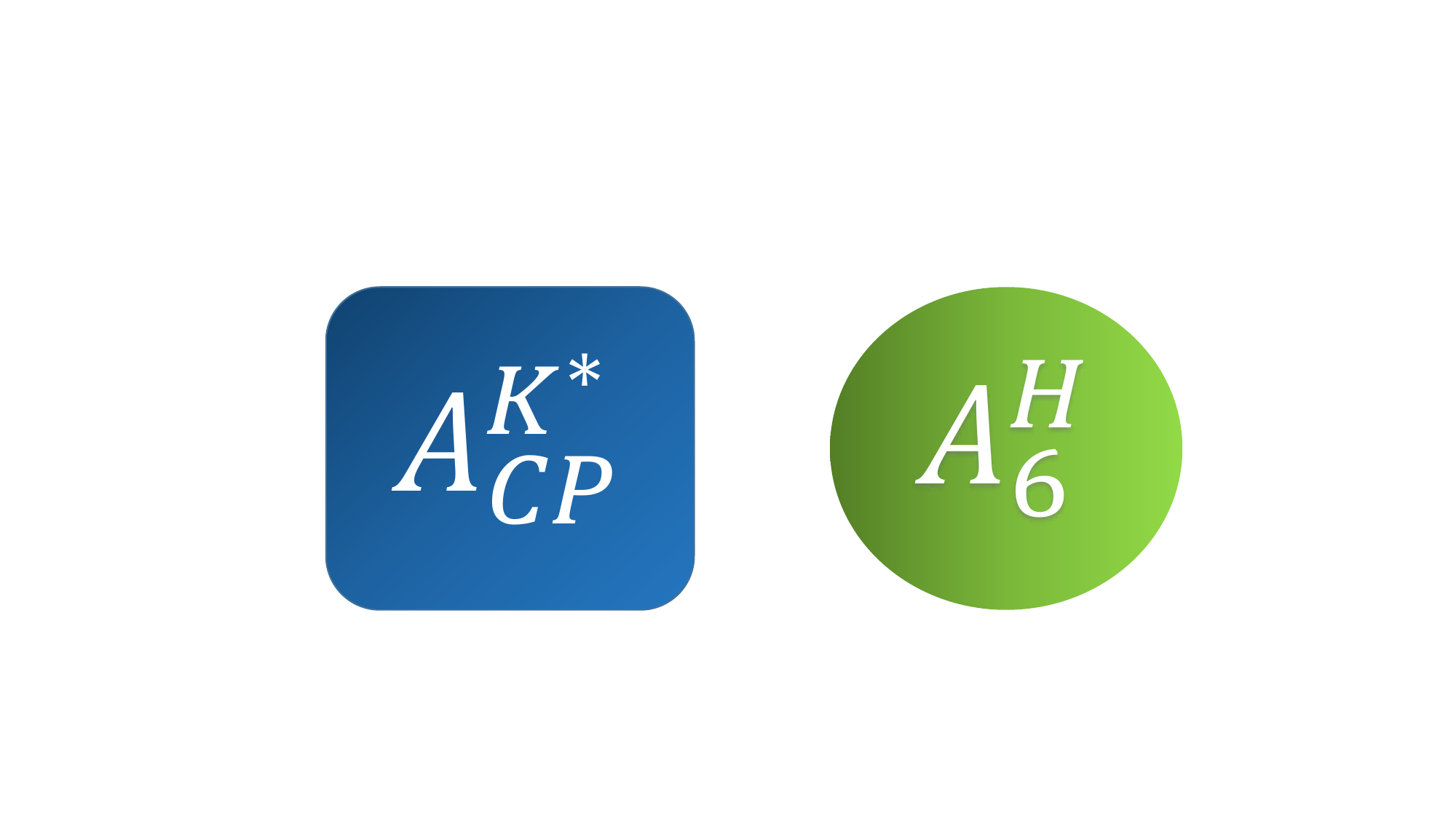}}
\newcommand{\imgtwo}{\includegraphics[scale=0.10]{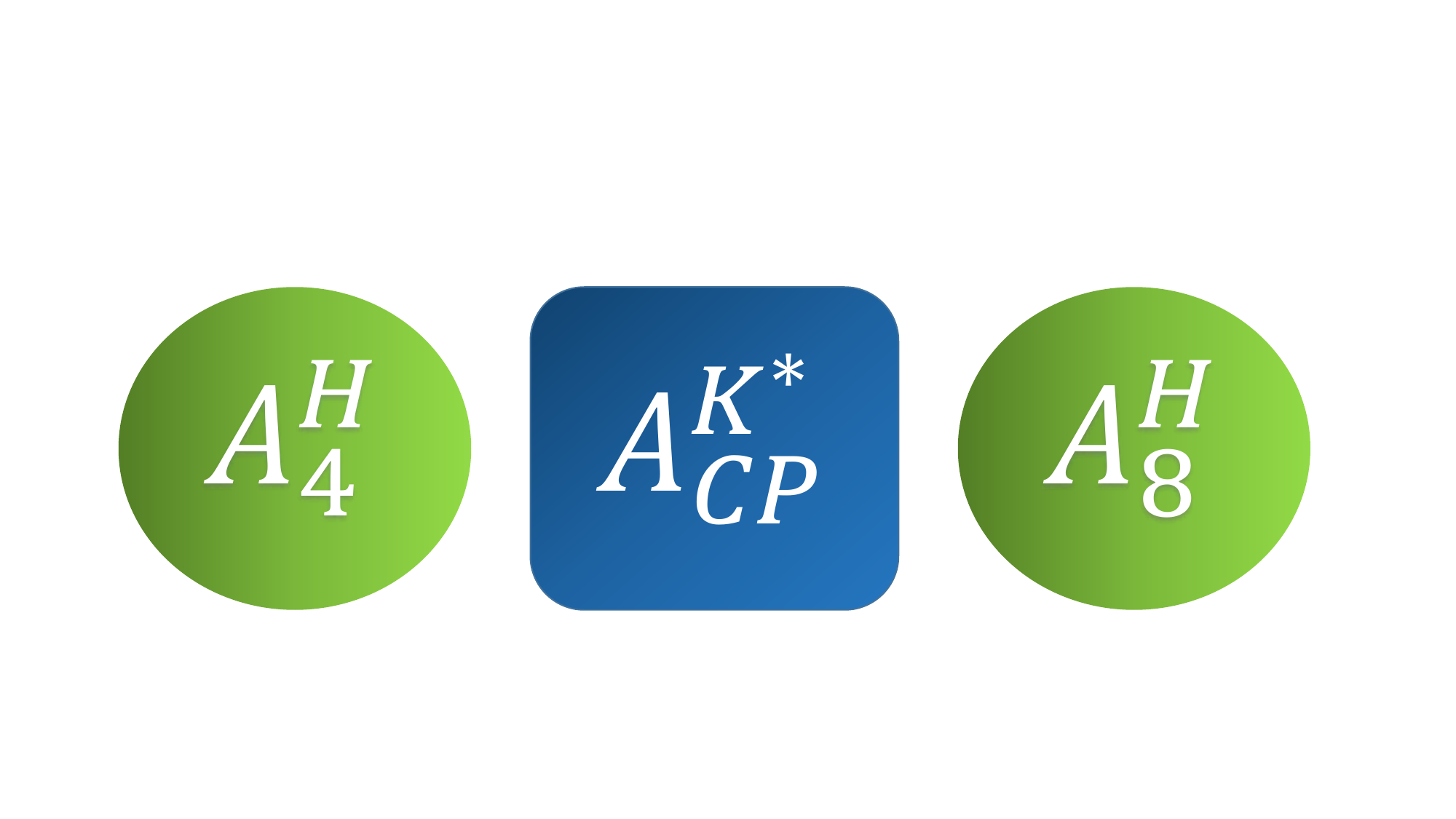}}
\newcommand{\imgthree}{\includegraphics[scale=0.10]{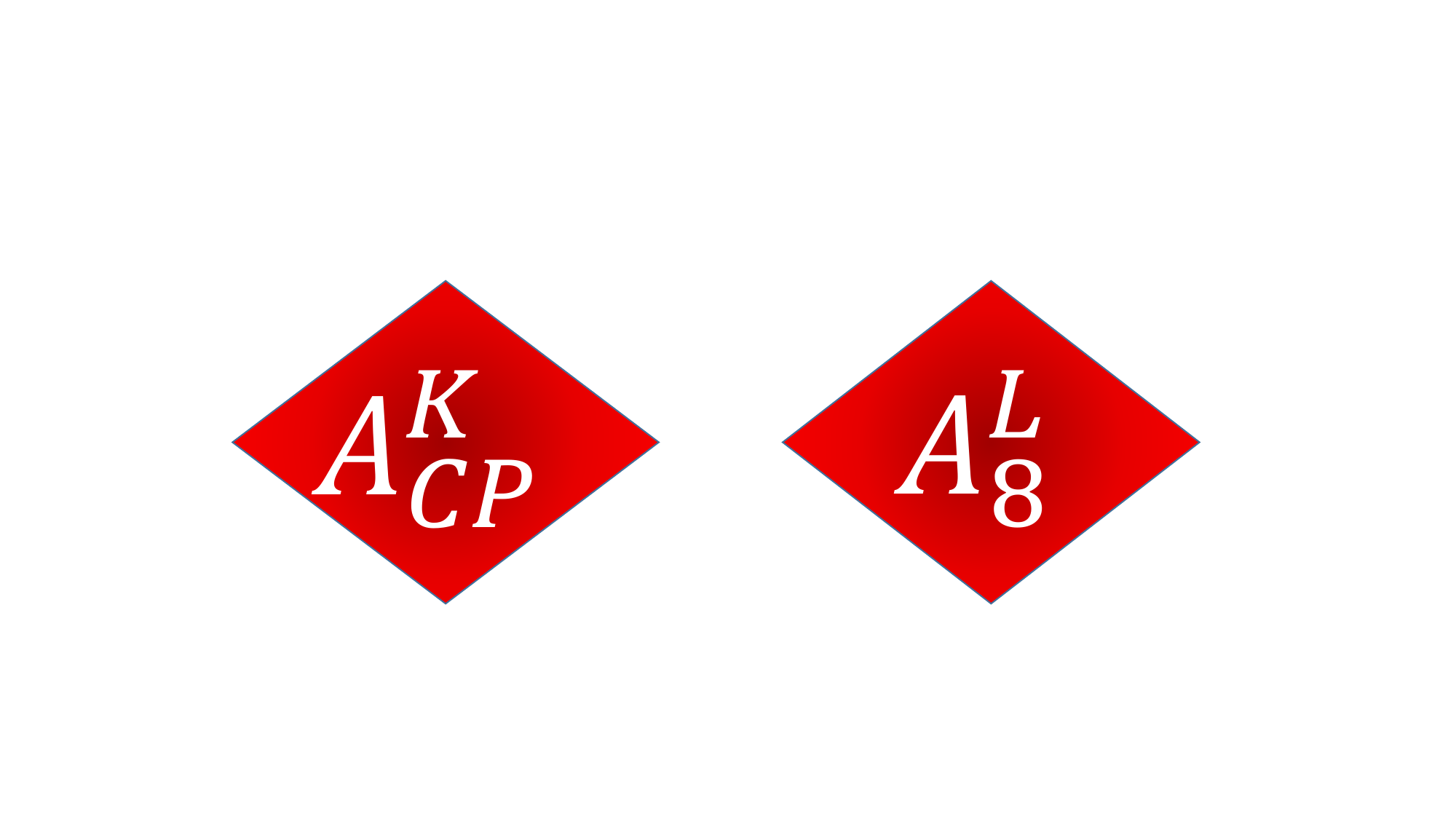}}
\newcommand{\imgfour}{\includegraphics[scale=0.10]{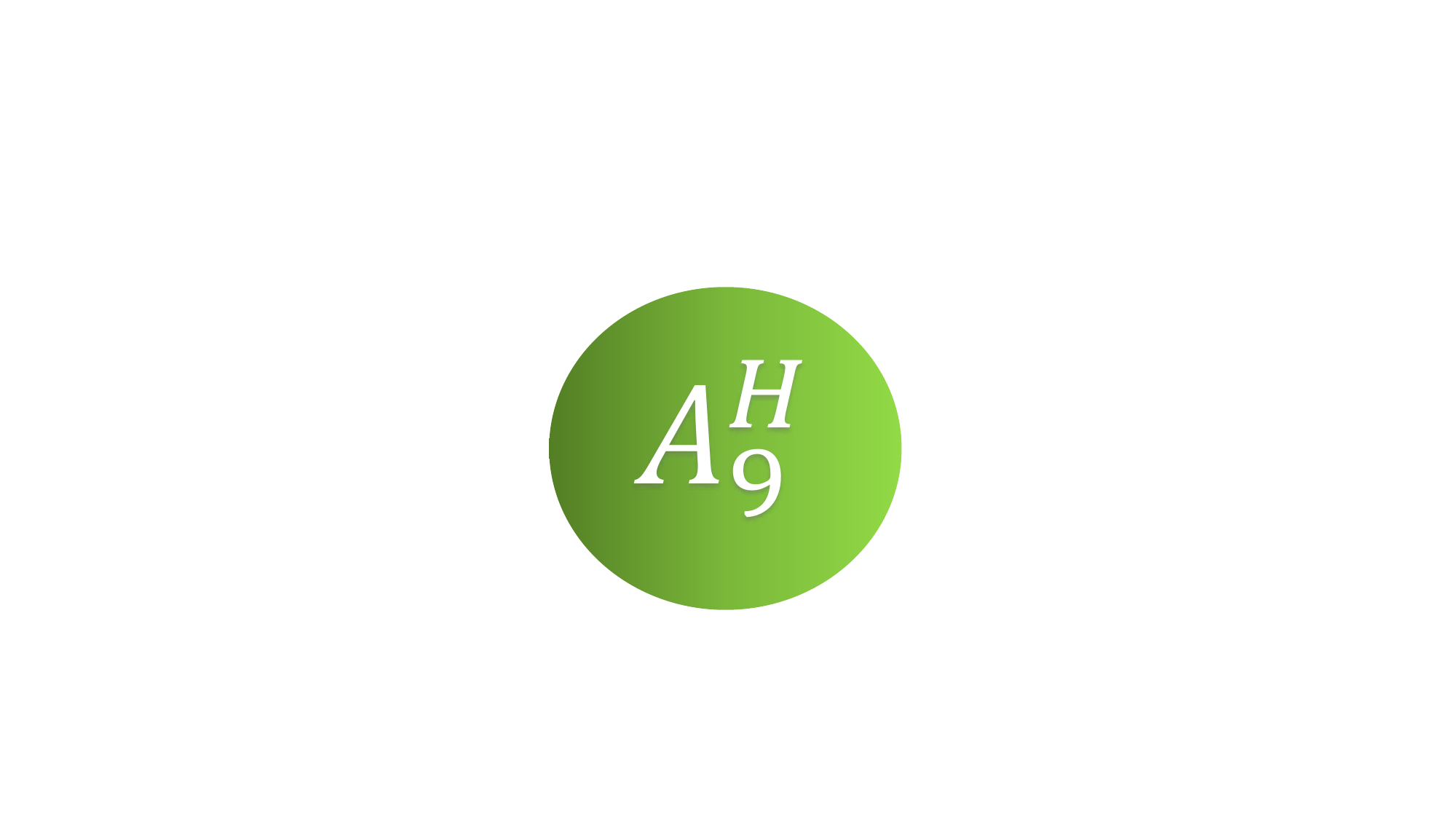}}
\newcommand{\imgfive}{\includegraphics[scale=0.10]{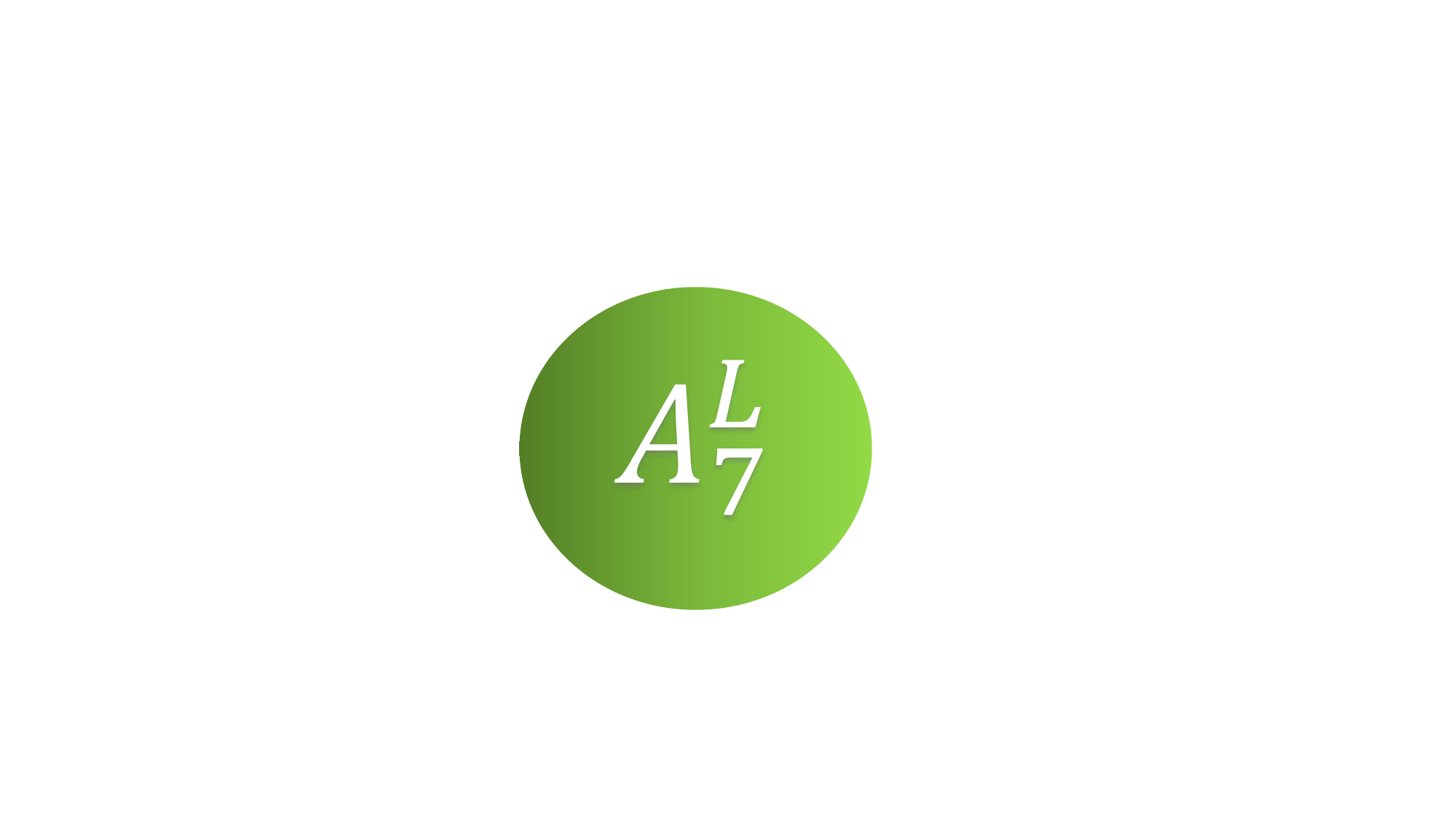}}
\begin{table*}[htb]
\begin{center}
\begin{tabular}{| p{3.4cm} | p{3.65cm} | p{3.65cm} | p{3.65cm}|}
\hline 
 \Large{1D Scenarios} & \hspace{.2cm} \Large{ $A^{\rm max} \approx$ (2-5)\% } &  \hspace{.2cm} \Large{$A^{\rm max} \approx$ (5-9)\%} &  \hspace{.3cm} \Large{$A^{\rm max} \gtrsim $ 10\%} \vspace{.3cm}\\    \hline  \hline
\hspace{1.3cm} \Large{$C_9^{NP}$} & \img &  &  \\
\hline
\hspace{.2cm}\Large{$C_9^{NP} = -C_{10}^{NP}$} & \imgone & \imgthree & \imgfive  \\
\hline
\hspace{.2cm}\Large{$C_9^{NP} = - C_{9}^{'}$} & \imgtwo & \imgfour & \\
\hline
\end{tabular}
\end{center}
\caption{ A plot exhibiting discriminating capabilities of various $CP$ violating observables in  $B^+ \to K^{+} \mu^+ \mu^-$ and $B^0 \to K^{*0} \mu^+ \mu^-$ decays. In each row, we show  new physics solutions along with the observables where a meaningful enhancement is allowed. These observables are further classified in three categories
on the basis of maximum amount of  enhancements, $A^{\rm max}$,  allowed by the current data. It is obvious that any observable which is placed in $A^{\rm max} \gtrsim $ 10\% or $A^{\rm max} \approx$ 5-9\% column also appears in the preceding columns. The observables marked in green color are termed as unique identifier of the specific new physics solution (appearing in the same row in which these observables appear), i.e., these observables will not appear in any other row. The observables marked in blue and red colors are degenerate observables in the sense that they appear in more than one row. Here and $A_{7,8}^L \equiv {A_{7,8}}_{[1-6]}$ and  $A_{4,6,8,9}^H \equiv {A_{4,6,8,9}}_{[15-19]}$. Further, $A^{K}_{\rm CP}$ and $A^{K^*}_{\rm CP}$ observables are in high-$q^2$ bin.  }
\label{summary}
\end{table*}

Apart from $A_{\rm CP}$, in this work we consider consider $A_{3,\,4,\,5}$, $A_6^s$ and $A_{7,\,8,\,9}$ observables. These observables are measured by the LHCb collaboration, however, with large errors \cite{Kstarlhcb2}. The angular observables $A_{3,\,4,\,5}$ and $A_6^s$ are direct $CP$ like asymmetries whereas $A_{7,\,8,\,9}$ are triple product CP asymmetries \cite{Bobeth:2008ij}. Therefore $A_{7,\,8,\,9}$ observables seem to be more sensitive to new weak phases as compared to the other observables.

First of all, we examine $A^{K^*}_{\rm CP}$. Based on predictions obtained in Table \ref{pred-acp1} for various favored scenarios, it is pellucid that none of the solutions can enhance
$A^{K^*}_{\rm CP}$ to a level of a percent in the low-$q^2$ bin. Therefore, the current $b \to s \ell \ell$ data suggests that the measurement of $A_{\rm CP}$ in $B^0 \to K^{*0} \mu^+ \mu^-$ decay in the low-$q^2$ region would be a hellacious task. On the contrary, in the high-$q^2$ bin, all favored new physics scenarios can ameliorate $A^{K^*}_{\rm CP}$ up to 2-3\%. However, as the maximum allowed value of $A^{K^*}_{{\rm CP} {[15-19]}}$ for all solutions are close to each other, one needs to look for additional $CP$ violating observables to discriminate between the allowed solutions.

The predictions of several $CP$ violating angular observables in $B^0 \to K^{*0} \mu^+ \mu^-$ in the low and high-$q^2$ regions are exhibited in Table \ref{pred-1} and \ref{pred-2}, respectively. From Table \ref{pred-1}, it is unambivalent that all three allowed solutions predict $A_{3,\,4,\,5}$, $A_6^s$ and $A_9$ asymmetries to be less than a percent in the low-$q^2$ bin and hence making their observation an arduous endeavor. However, the prediction of observable $A_7$ provides encouraging sign  for $C_9^{NP} = -C_{10}^{NP}$ solution for which ${A_7}_{[1-6]}$ can be enhanced up to 10\%. For all other solutions,  ${A_7}_{[1-6]}<1\%$. Therefore the measurement of ${A_7}_{[1-6]}$  observable can lead to a unique identification of  new physics solution  in the form of $C_9^{NP} = -C_{10}^{NP}$. 
The $C_9^{NP}$ and $C_9^{NP} = -C_{10}^{NP}$ solutions can bolster ${A_8}_{[1-6]}$ at the level of  4-5\% whereas for $C_9^{NP} = - C_{9}^{'}$ scenario, ${A_8}_{[1-6]} \lesssim 1 \%$. Therefore measurement of ${A_8}_{[1-6]}$ at the level of few percent would discriminate  $C_9^{NP} = - C_{9}^{'}$ new physics scenario from other two scenarios. As both $C_9^{NP}$ and $C_9^{NP} = -C_{10}^{NP}$ solutions allow almost similar enhancement in the value of ${A_8}_{[1-6]}$, a discrimination between these two solutions would not be possible through this observable. 

\begin{figure*}[htb]
\begin{center}
\includegraphics[width=0.45\textwidth]{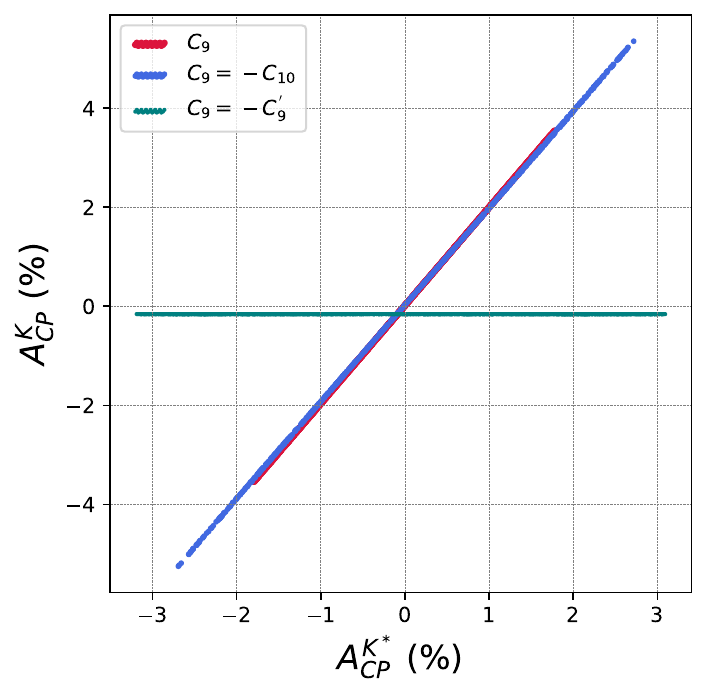} 
\includegraphics[width=0.45\textwidth]{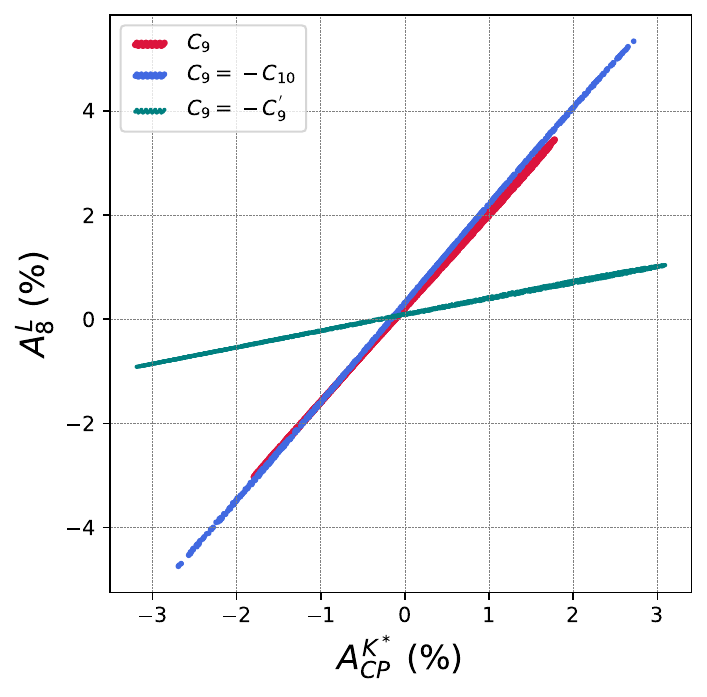}
\end{center}
\caption{The left panel portrays correlations between $A_{\rm CP}$ in   $B^0 \to K^{*0} \mu^+ \mu^-$ and  $B^+ \to K^{+} \mu^+ \mu^-$ in the high-$q^2$ region for  all favored ``1D" solutions. A correlation between $A_{\rm CP}$ in $B^0 \to K^{*0} \mu^+ \mu^-$ in the high-$q^2$ bin and $A_8$ observable in the low-$q^2$ region ($A_8^L$) is depicted in the right panel.}
\label{corr-acp-0}
\end{figure*}

\begin{figure*}[htb]
\begin{center}
\includegraphics[width=0.45\textwidth]{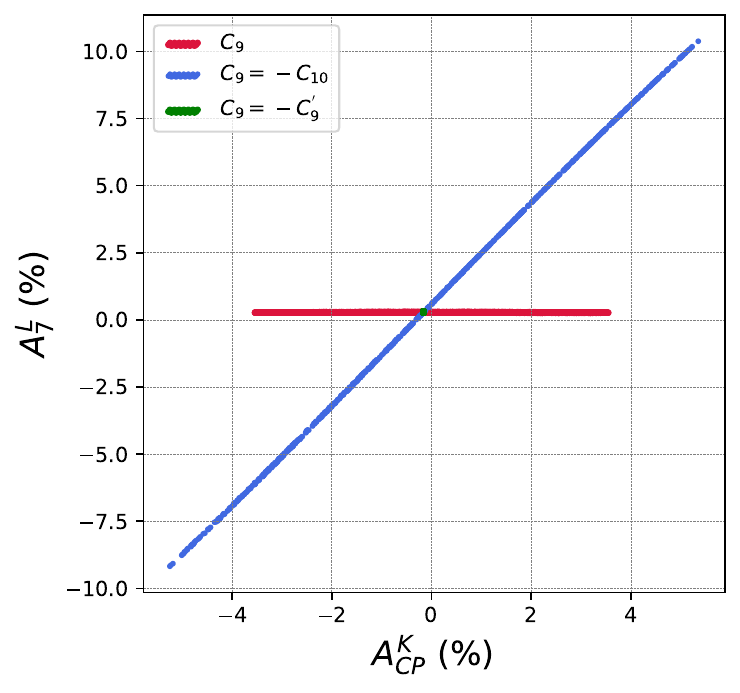}
\includegraphics[width=0.43\textwidth]{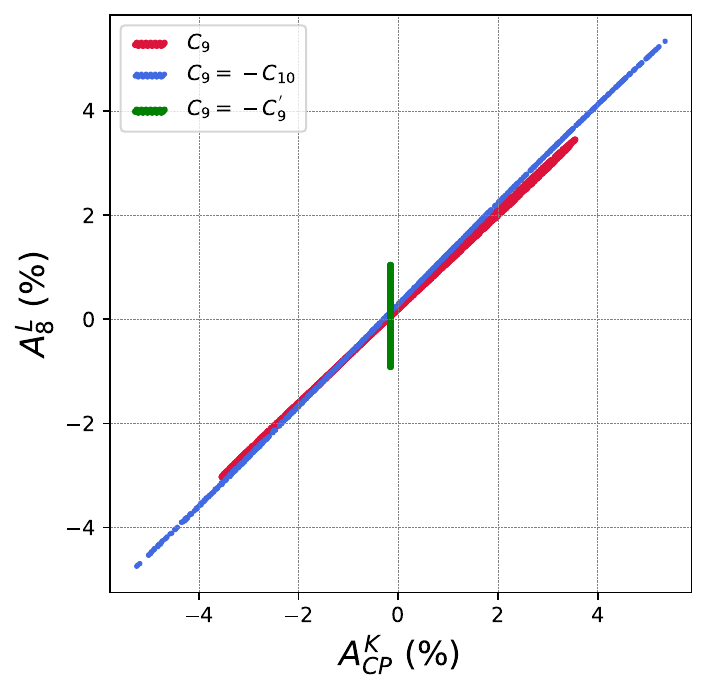}
\end{center}
\caption{The left panel portrays correlations between $A_{\rm CP}$ in $B^+ \to K^{+} \mu^+ \mu^-$ in the high-$q^2$ and $A_7$ observable in the low-$q^2$  region for  all favored ``1D" solutions. A correlation between $A_{\rm CP}$ in $B^+ \to K^{+} \mu^+ \mu^-$ in the high-$q^2$ bin and $A_8$ observable in the low-$q^2$ region is depicted in the right panel.}
\label{corr-a7a8-0}
\end{figure*}

\begin{figure*}[htb]
\begin{center}
\includegraphics[width=0.45\textwidth]{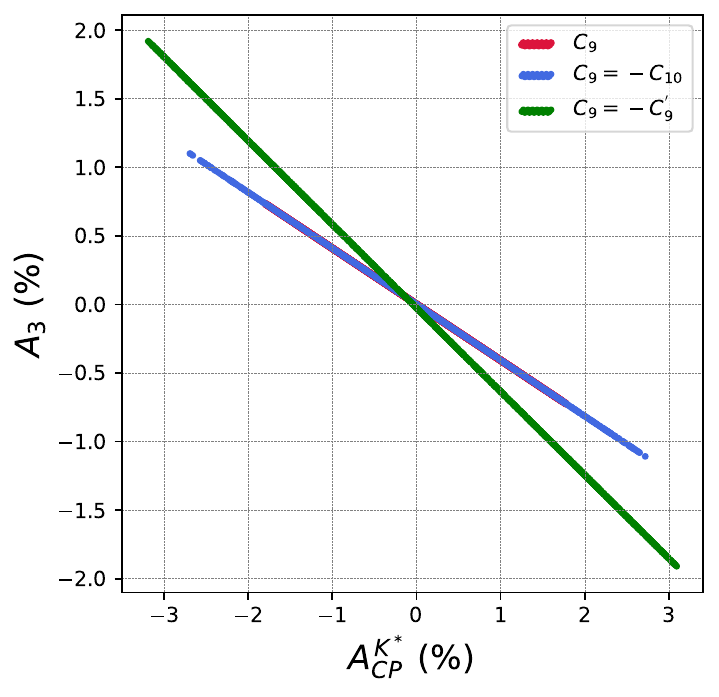} 
\includegraphics[width=0.45\textwidth]{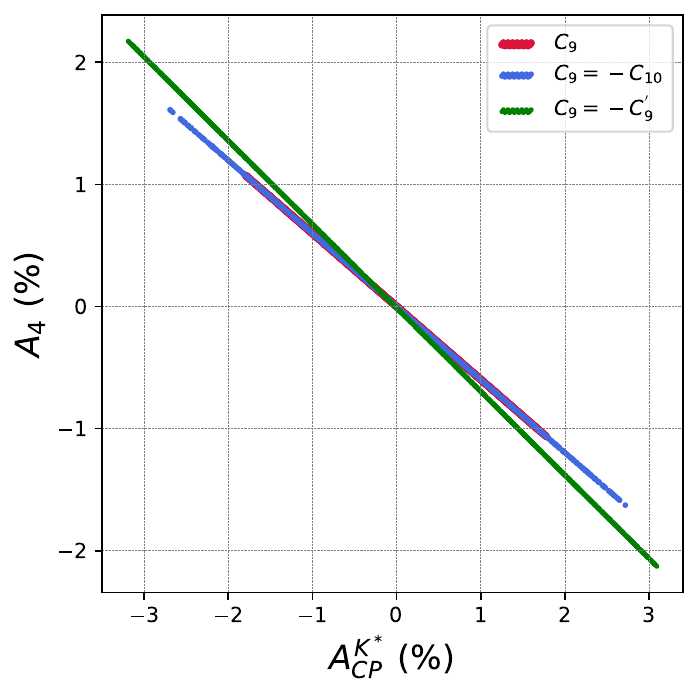}\\
\includegraphics[width=0.45\textwidth]{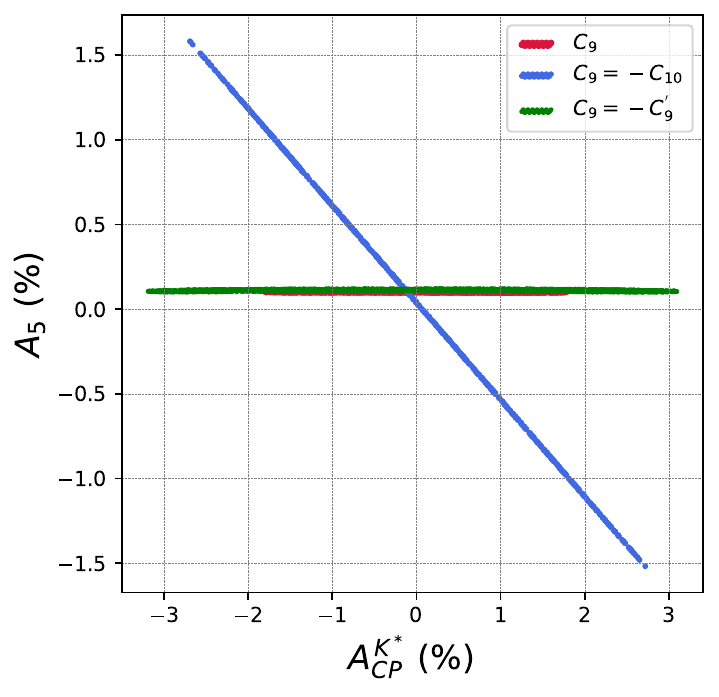} 
\includegraphics[width=0.45\textwidth]{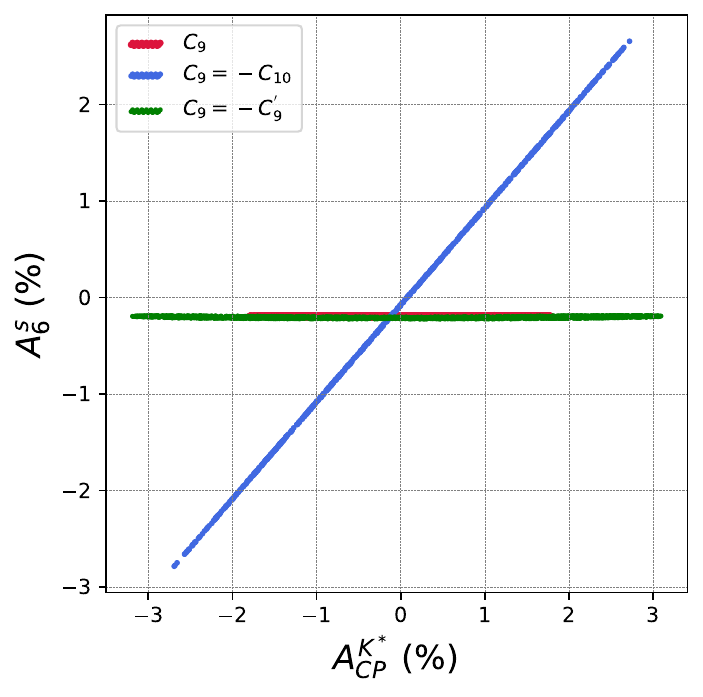}\\
\includegraphics[width=0.45\textwidth]{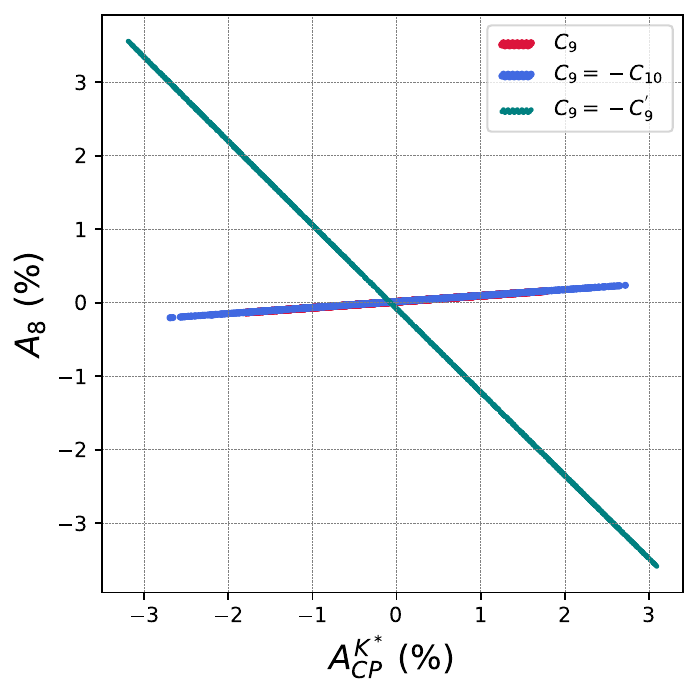} 
\includegraphics[width=0.45\textwidth]{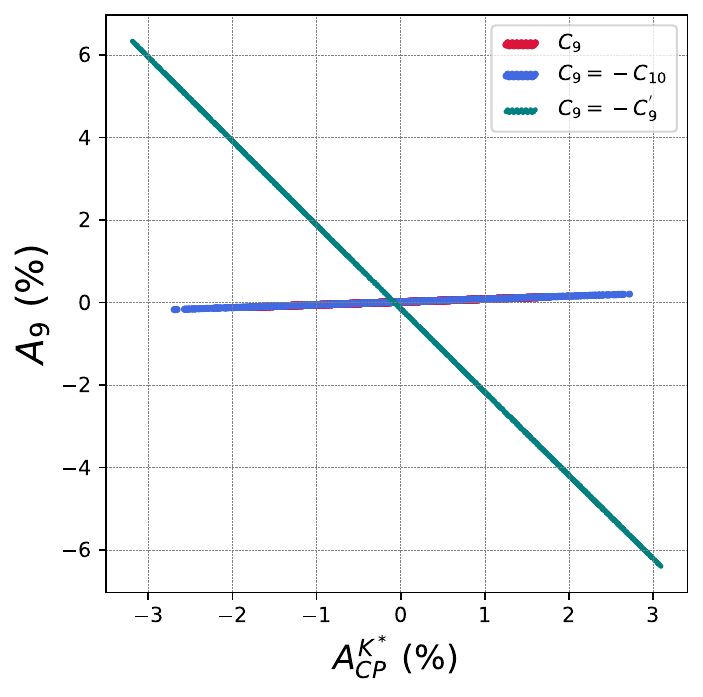}
\end{center}
\caption{These plots  reveal correlations between $A_{\rm CP}$ in   $B^0 \to K^{*0} \mu^+ \mu^-$ and several $A_i$ observables for  all favored ``1D" solutions in the high-$q^2$ bin.}
\label{corr-acp-high}
\end{figure*}

We now consider predictions of $A_i$ observables in the high-$q^2$ bin as given in Table \ref{pred-2}. The most conspicuous feature of predictions in the low-$q^2$ region was related to the observable $A_7$ which insinuated to be a potential observable to verbalize the signatures of weak phase related to the new physics solution $C_9^{NP} = -C_{10}^{NP}$. However, unlike in the low-$q^2$ bin, all allowed 1D solutions fail to provide an enhancement in $A_7$ above a percent level in the high-$q^2$ region. The $C_9^{NP} = - C_{9}^{'}$ scenario failed to make any noticeable indentations  in the low-$q^2$ bin as it was unable to provide any detectable enhancements in any of the considered $A_i$ observables. However, in the high-$q^2$ bin, this solution appears to make a riveting impact as it can enhance $A_8$ and $A_9$ observables up to a level $\sim $ (4 - 6)\%. All other favored scenarios fail to provide any meaningful enhancement in these observables. This thus implies that the observation of either $A_8$ or $A_9$ asymmetries at the level of a few percent in the high-$q^2$ bin may provide confirmatory evidence in support of the $C_9^{NP} = - C_{9}^{'}$ scenario.

The observable $A_6^s$ failed to make any imprint in the low-$q^2$ bin. However in [15-19] bin, this can be enhanced up to $\sim$ (2 - 3)\% by  $C_9^{NP} = -C_{10}^{NP}$ scenario. The other two scenarios predict $A_6^s < 1\%$. Therefore the observation of  $A_6^s$  in  the high-$q^2$  bin at the level of a few percent  would provide an unambiguous signature of new physics in the form of $C_9^{NP} = -C_{10}^{NP}$ solution.  None of the  1D solutions  in [1-6] bin provided any meaningful enhancements in $A_{3,\,4,\,5}$ angular observables. On the contrary, all of these observables can be enhanced up to a level of a few percent in the high-$q^2$ bin. The $A_5$ observable can be brought to a percent level  by  $C_9^{NP} = -C_{10}^{NP}$ solution. All solutions   have the potential to bring $A_4$ observable up to a level of a percent or more. A similar job for ${A_3}_{[15-19]}$ observable can be done by $C_9^{NP} = -C_{10}^{NP}$ and $C_9^{NP} = - C_{9}^{'}$ solutions.

Thus we see that an accurate measurement  of $A_{CP}$ in the high-$q^2$ bin along with a number of $CP$ violating angular observables would enable unique identification of possible new physics in $b \to s \ell \ell$ transition. This can be easily understood with the help of summary Tab.~\ref{summary}. In this table,   we have listed those observables   for which the current allowed solutions can provide meaningful enhancement, say values above 2\%. For each new physics solutions, the observables are distributed in different blocks based on their allowed values. 

The observables represented in green color are unique identifiers. For e.g, the observable ${A_{7}}_{[1-6]}$ appears in all three blocks. This implies that the measurement of ${A_{7}}_{[1-6]}$ with any value greater than 2\% can provide confirmatory evidence for new physics in the form of $C_9^{NP} = -C_{10}^{NP}$ solution. Similarly, measurement of $A_6^s$  in (2-5)\% range would turn out to be another unique identifier of $C_9^{NP} = -C_{10}^{NP}$ scenario. For $C_9^{NP} = - C_{9}^{'}$ solution, $A_{4,8,9}$ observables in the high-$q^2$ bin will serve this task. Further, the $C_9^{NP} $ solution  
doesn't have any unique identifier, i.e. its confirmation would require  measurements of more than one observable. 

The observables represented by blue and red colors provide signatures of new physics for multiple scenarios.  The $A_{\rm CP}^{K^*}$ in the high-$q^2$ bin is allowed to have values in the range of $(2-5)$\% for all three scenarios. Similarly, $A_{\rm CP}^{K}$ in high-$q^2$ and $A_8$ in low-$q^2$ bin can be enhanced by $C_9^{NP} $ as well as  $C_9^{NP} = -C_{10}^{NP}$ scenarios. Therefore, a careful scrutinization is required to see whether these observables in combination with others can also provide unique identification. It is apparent from the table that the following combinations can also serve as a useful discriminant for new physics solutions:

\begin{itemize}

\item {[$A_{\rm CP}^{K}$ - $A_{\rm CP}^{K^*}$] in high-$q^2$:} A simultaneous measurement of these observables  can be used as a good discriminant for $C_9^{NP} = - C_{9}^{'}$ solution. The $C_9^{NP} $ as well as $C_9^{NP} = -C_{10}^{NP}$ solutions allow both of these observables to have values greater than 2\% whereas the $C_9^{NP} = - C_{9}^{'}$ scenario doesn't allow any meaningful enhancement in $A_{\rm CP}^{K}$. Therefore any  measurement of both $A_{\rm CP}^{K}$ and $A_{\rm CP}^{K^*}$ in high-$q^2$ region at the level of few percent would disfavor $C_9^{NP} = - C_{9}^{'}$ scenario. The same is corroborated from the correlation plot between these observables as depicted in the left panel of Fig.~\ref{corr-acp-0}. It is obvious that both $A_{\rm CP}^{K}$ and  $A_{\rm CP}^{K^*}$ enjoy positive correlations with each other for $C_9^{NP} $ and $C_9^{NP} = -C_{10}^{NP}$ solutions, i.e., a finite measurement of one observable would imply the same for the other whereas for $C_9^{NP} = - C_{9}^{'}$ scenario, a finite value of $A_{\rm CP}^{K^*}$  implies $A_{\rm CP}^{K} \approx 0$ in the entire high-$q^2$ bin.

\item ${A_{8}}_{[1-6]}$ - $A_{\rm CP}^{K^*}$ (high-$q^2$): A simultaneous measurement of these observables  can discriminate between $C_9^{NP} = - C_{9}^{'}$ solution with others. The other two solutions allow meaningful enhancements in both of these observables whereas  $C_9^{NP} = - C_{9}^{'}$ solution can boost only $A_{\rm CP}^{K^*}$ in the high-$q^2$ bin. This feature is also reflected from the right panel of Fig.~\ref{corr-acp-0}.

\end{itemize} 

For a finer scrutinization of possible new  physics contributions to $b \to s \ell \ell$, we now investigate other correlations between $A_{\rm CP}$ and $CP$ violating angular observables $A_i$'s in $B^0 \to K^{*0} \mu^+ \mu^-$. We first examine these correlations between $A_{\rm CP}^{K}$ in the high-$q^2$ bin and ${A_{7,8}}_{[1-6]}$ as for all other angular observables in the low-$q^2$ region, enhancements are too small to be observed. Further, we do not consider ${A_{7,8}}_{[1-6]}$ correlations with $A_{\rm CP}^{K}$ in the low-$q^2$ bin as  $A_{\rm CP}^{K}<1\%$ in this bin.
 These correlations are vignetted in Fig.~\ref{corr-a7a8-0}.
 For $C_9^{NP} = -C_{10}^{NP}$ solution,  $A_{\rm CP}^{K}$  and ${A_{7}}_{[1-6]}$ observables are correlated in the sense that the larger enhancement in  $A_{\rm CP}^{K}$  will commensurate a larger enhancement in ${A_{7}}_{[1-6]}$ as well. For e.g., for $A_{\rm CP}^{K} \approx 5\%$, ${A_{7}}_{[1-6]} $ can be  $\sim 10\%$. For $C_9^{NP} $  solution, ${A_{7}}_{[1-6]} < 1\%$ for  the entire allowed range of $A_{\rm CP}^{K}$  and both of these observables remain unstirred for the $C_9^{NP} = - C_{9}^{'}$ solution.

The correlations between $A_{\rm CP}^{K}$ in the high-$q^2$ bin and ${A_{8}}_{[1-6]}$ are explicated in the right panel of Fig.~\ref{corr-a7a8-0}.
 For  $C_9^{NP} = -C_{10}^{NP}$ scenario, the maximum possible enhancement in $A_{\rm CP}^{K}\approx 5\%$ can be accustomed with the maximum allowed value of ${A_{8}}_{[1-6]}\approx 5\%$ and vice-versa. For $C_9^{NP} $ scenario, ${A_{8}}_{[1-6]}$ and $A_{\rm CP}^{K}$ have positive correlations and their behavior is identical to that of  $C_9^{NP} = -C_{10}^{NP}$ solution but with slightly smaller enhancement ($\approx 3\%$ ) in both   observables. The $A_{\rm CP}^{K}$  remains unaltered for the $C_9^{NP} = - C_{9}^{'}$ solution with $\leq 1 \%$ enhancement in ${A_{8}}_{[1-6]}$ observable.

We now delve correlations in the high-$q^2$ region.  The interrelations between $A_{\rm CP}^{K^*}$ and $A_i$  are demonstrated in Fig.~\ref{corr-acp-high}. Here we do not consider $A_7$ observable as none of the allowed 1D scenarios can enhance it up to a level of a percent. The $(A_{\rm CP}^{K^*} - A_3)$ and $(A_{\rm CP}^{K^*}-A_4)$ correlations are almost similar. These $CP$ violating angular observables are anti-correlated with $A_{\rm CP}^{K^*}$ for the three allowed solutions, i.e., a negative value of $A_{\rm CP}^{K^*}$ would imply $A_{3,4} >0$ and vice versa. Further,   $|A_{\rm CP}^{K^*}|\approx 3\%$, which is the maximum allowed value of $A_{\rm CP}^{K^*}$ with the current data, can lead to $|A_{3,4}| \approx 2\%$  for $C_9^{NP} = - C_{9}^{'}$ scenario. For $C_9^{NP}$ and $C_9^{NP} = -C_{10}^{NP}$ scenarios, $A_{\rm CP}^{{K^*}{\rm max}}$ implies $A_{3,4} \approx 1\%$.

  From $(A_{\rm CP}^{K^*}-A_{5,6}^{(s)})$ plots, it is obvious that  $A_{5}$ ($A_6^{s}$) has  negative (positive) correlations with $A_{\rm CP}^{K^*}$ for $C_9^{NP} = -C_{10}^{NP}$ solution. For e.g., a measurement of $A_{\rm CP}^{K^*}$ with a value $\approx -3\%$ would lead to an observation of $A_5$ ($A^s_6$) with a value $\approx 1\%$  ($\approx -2\%$). Therefore  simultaneous measurements of $A_{\rm CP}^{K^*}$ and $A_{5,6}^{(s)}$ can discriminate $C_9^{NP} = -C_{10}^{NP}$ solution from others.  The $(A_{\rm CP}^{K^*}-A_8)$ and $(A_{\rm CP}^{K^*}-A_9)$ correlations features are almost similar for the three scenarios.  Here $A_{8,9}$ have anti-correlations with $A_{\rm CP}^{K^*}$ for the $C_9^{NP} = - C_{9}^{'}$ scenario. For this solution,  a measurement of   $A_{\rm CP}^{K^*} \approx -3\%$ would imply $A_{8,9}\approx 5\%$.

As of now, we have been emphasizing on the measurements of the angular observables $A_i$ along with   $A_{\rm CP}$'s for discriminating between the allowed solutions. However, 
these correlations will also be helpful in discarding or identifying a particular scenario with a precise measurement of  $A_{\rm CP}$ even if we only have upper bounds on the $A_i$ observables. For e.g, from $(A_{\rm CP}^{K^*} - A_8)$ correlation plot in the high-$q^2$ region, it is evident that a finite value of $A_{\rm CP}^{K^*}$ indicates a finite value of $A_8$ for $C_9^{NP} = - C_{9}^{'}$ scenario. In case, the experimental upper bounds on $A_8$ slips below the value predicted by the correlation plot (on the basis of the measured value of $A_{\rm CP}^{K^*}$), the given scenario would be disfavored.

This may also help in the  identification of $C_9^{NP}$ solution. As evident from the left panel of  Fig.~\ref{corr-acp-0}, for this solution (as well as $C_9^{NP} = -C_{10}^{NP}$), $A_{\rm CP}^{K}$ and  $A_{\rm CP}^{K^*}$ in the high-$q^2$  have positive correlation, i.e a larger value in one will imply the same for other observable. Therefore if both of these observables are measured, say with a value $\gtrsim 2\%$, this can only be due to either $C_9^{NP}$ or $C_9^{NP} = -C_{10}^{NP}$ scenario. This degeneracy can be removed by inspecting correlations of $A_{\rm CP}$ with $A_i$ observables. For e.g.,  ($A_{\rm CP}^{K^*}$ - $A_6^s$) plot predicts    $|A_6^s| \approx 2\%$ for $C_9^{NP} = -C_{10}^{NP}$   solution corresponding to $|A_{\rm CP}^{K^*}|\approx 2\%$ and $\approx 0$ for $C_9^{NP}$ scenario. Therefore if the experimental upper bound on $|A_6^s|$ observable falls below 2\%,  such a scenario can only be accommodated by the $C_9^{NP} $ solution. 

On similar lines, the simultaneous measurements of $A_{\rm CP}^{K}$ and  $A_{\rm CP}^{K^*}$ in the high-$q^2$ bins can be used as a good identifier between the possible NP solutions. In case, $A_{\rm CP}^{K^*}$ is measured at the level of $\gtrsim 2\%$ and the upper bounds on $A_{\rm CP}^{K}$ shrinks  to less than $2\%$, it would be enough to identify $C_9^{NP} = - C_{9}^{'}$ solution by disfavoring the other two.

\section{Conclusions}
\label{conc}
Assuming new physics Wilson coefficients to be complex, we perform a model-independent global fit to all apropos $b \to s \ell \ell$ ($\ell=e,\,\mu$) data. This include updated  measurements of $R_{K}$ and $R_{K^{*}}$ by the LHCb collaboration in December 2022 together with the updated measurement of the branching ratio of $B_s \to \mu^+\mu^- $ by the CMS collaboration and the measurements of several $B_s \to \phi \mu^+\mu^-$ observables. We work under the assumption that the new physics equally affects both  the muon and electron sectors. For comparison, we also update the fits for real couplings under the assumption of universal couplings.  Considering only one operator or two related operators at a time, we obtain the following:

\begin{itemize}

\item The allowed solutions remain the same as obtained for the real fits, i.e. $C_9^{\rm NP}$, $C_9^{\rm NP} = -C_{10}^{\rm NP}$ and $C_9^{NP} = - C_{9}^{'}$ scenarios still provide a good or moderate fits to the data.

\item The $C_9^{\rm NP}$ and $C_9^{NP} = - C_{9}^{'}$  scenarios now becomes the most preferred one as the $\Delta \chi^2 $ for $C_9^{\rm NP} = -C_{10}^{\rm NP}$ solution falls by $\sim$ 10 below $\Delta \chi^2 $ values for the other two scenarios. Therefore 
the $C_9^{\rm NP} = -C_{10}^{\rm NP}$  scenario can only be considered as a moderate solution.

\item The $C_{10}^{\rm NP}$  scenario which provided a moderate fit to the data before CMS and December 2022 LHCb updates now fails to provide any significant improvement in the value of $\Delta \chi^2 $.

\end{itemize}

We find that the current data allows complex couplings to exist with an upper bound similar to that of their real counterparts. The effect of such a weak phase can show up in some of the $CP$ asymmetries.   For the favoured solutions, we obtain predictions of several $CP$-violating observables in $B^0 \to K^{*0} \mu^+ \mu^-$ and direct $CP$ asymmetry in $B^+ \to K^+ \mu^+ \mu^-$. These asymmetries can be observed at the current or planned experimental facilities provided new physics enhances them up to a level of a few percent. Following are our main observations:

\begin{itemize}

\item None of the new physics solutions can enhance $A_{\rm CP}$ in the low-$q^2$ bin at the level of a few percent. Such an enhancement is feasible only in the high-$q^2$ region. This is true for $B^+ \to K^+ \mu^+ \mu^-$ as well as $B^0 \to K^{*0} \mu^+ \mu^-$ decay. For $B^+ \to K^+ \mu^+ \mu^-$, such an enhancement can be provided by $C_9^{NP}$ or $C_9^{NP} = -C_{10}^{NP}$ solutions, the enhancement being more pronounced for the later solution. For $B^0 \to K^{*0} \mu^+ \mu^-$ decay, all solutions can serve this purpose.

\item  All allowed solutions predict $A_{3,\,4,\,5}$, $A_6^s$ and $A_9$ asymmetries to be less than a percent level in the low-$q^2$ bin. However, the predictions of observable $A_7$ provides eupeptic sign for $C_9^{NP} = -C_{10}^{NP}$ solution as ${A_7}_{[1-6]}$ can be enhanced up to 10\%. Therefore ${A_7}_{[1-6]}$ can be termed as an unique identifier for $C_9^{NP} = -C_{10}^{NP}$ solution.

\item The $C_9^{NP}$ and $C_9^{NP} = -C_{10}^{NP}$ solutions can bolster ${A_8}_{[1-6]}$ at the level of  4-5\%.

\item  The observation of any of the $A_4$, $A_8$ or $A_9$ observables at a level of a few percent in the high-$q^2$ bin may provide confirmatory evidence in support of the $C_9^{NP} = - C_{9}^{'}$ scenario.

\item The observable $A_{3,\,4,\,5}$ and $A_6^s$ failed to make any impact in the low-$q^2$ bin. However in [15-19] bin,  all of these observables can be enhanced up to a level of a  percent or more. A measurement of $A_6^s$ up to (2-3)\% level would provide unique identification of  $C_9^{NP} = -C_{10}^{NP}$ solution.

\end{itemize}

Finally, we study correlations between $A_{\rm CP}$ and other $CP$ asymmetries. Our findings are as follows:

\begin{itemize}

\item A simultaneous measurement of $A_{\rm CP}^{K}$ and $A_{\rm CP}^{K^*}$ in the high-$q^2$  bin can be used as a good discriminant for $C_9^{NP} = - C_{9}^{'}$ solution. The same can also be achieved  by simultaneous measurements of ${A_{8}}_{[1-6]}$ and $A_{\rm CP}^{K^*}$ in the high-$q^2$ region. 

\item The $(A_{\rm CP}^{K^*}-A_3)$ and $(A_{\rm CP}^{K^*}-A_4)$ correlations in the high-$q^2$ bin cannot discriminate between any of the solutions whereas a simultaneous measurement of $A_{\rm CP}^{K^*}$ and  $A_5$ (or $A_6^s$) in high-$q^2$ region can distinguish $C_9^{NP} = -C_{10}^{NP}$ from other scenarios. A similar identification for $C_9^{NP} = - C_{9}^{'}$ solution can be provided by examining $(A_{\rm CP}^{K^*}-A_{8,9})$ correlations in the high-$q^2$ bin.

\item If $A_{\rm CP}$ is precisely measured in the high-$q^2$ region, the new physics solutions can  also be identified even if we only have upper bounds on the $A_i$ observables.   For some scenarios, a discrimination would be possible only through ($A_{\rm CP}^{K}$ - $A_{\rm CP}^{K^*}$) correlations in the high-$q^2$ bin.

\end{itemize}
Therefore the observation  of  $A_{\rm CP}$ as well as $CP$ violating angular observables will not only provide an evidence of new physics with complex phase but their accurate measurements would also facilitate the unique identification of possible new physics in the decays induced by the $b \to s \ell \ell$ transition. The direct $CP$  asymmetry can be measured at the LHCb or Belle-II, however the measurements of $CP$ violating angular observables require higher statistics which can be attained  at the HL-LHC \cite{Cerri:2018ypt}.

 \bigskip
\noindent
{\bf Acknowledgements}: I would like to thank Roman Zwicky  for useful suggestions. I would also like to thank Arindam Mandal and Ashutosh Kumar Alok for helpful discussions on various aspects of the draft.

\appendix
\section{Decay rate of $B \to K \mu^+ \mu^-$ decay}
\label{appen0}
The decay rate of $B \to K \mu^+ \mu^-$ is given by \cite{Becirevic:2012fy,Bobeth:2007dw}
\begin{equation}
\Gamma (B \to K \mu^+ \mu^-) = \int_{q^2_{\rm min}}^{{q^2_{max}}} dq^2 \left(2a_{\mu}(q^2) +\frac{2}{3} c_{\mu}(q^2)\right)\,,
\end{equation}
where
\begin{eqnarray}
a_{\mu}(q^2) &=& E(q^2) \Bigg[ q^2 |F_P|^2 + \frac{\lambda}{4} \left( |F_V|^2 + |F_A|^2\right)\nonumber\\
&&+ 2m_{\mu} \left( m_B^2 -m_K^2 +q^2\right){\rm Re} (F_P F_A^{*}) \nonumber\\
&& + 4 m_{\mu}^2 m_B^2 |F_A|^2  \Bigg]\,,\\
c_{\mu}(q^2) &=& -\frac{\lambda}{4} \beta_{\mu}^2 E(q^2) \left( |F_V|^2 + |F_A|^2\right)\,,
\end{eqnarray}
with
\begin{equation}
E(q^2) = \frac{G_F^2\alpha |V_{tb} V_{ts}^*|^2}{512 \pi^5 m_B^3}\beta_{\mu}\lambda_K.
\end{equation}
Here $\lambda = m^4_B+m^4_{K}+q^4-2(m^2_B m^2_{K} +m^2_{B}q^2+m^{2}_{K}q^2)$ and $\beta_{\mu}= \sqrt{1-4m^2_{\mu}/q^2}$. In the low-$q^2$ region, all form-factors reduce to one soft form-factor \cite{Charles:1998dr,Beneke:2000wa}. In the high-$q^2$ region too, symmetry relations among the form factors can be delved with the improved Isgur-Wise relation \cite{Bobeth:2011nj}.

\section{Angular coefficients in $B^0 \to K^{*0} \mu^+ \mu^-$ decay}
\label{appen}

The angular coefficients appearing in the four-fold distribution of $B\to K^*(\to K\pi)\mu^+\mu^-$ decay  can be expressed in terms of transversity amplitudes as~\cite{Altmannshofer:2008dz}
\begin{widetext}
\begin{eqnarray}
I_1^s &=& \frac{(2+\beta^2_{\mu})}{4}\left[|A^L_{\perp}|^2+|A^L_{\parallel}|^2 +(L\to R)\right] \nonumber\\
&&+ \frac{4m^2_{\mu}}{q^2} {\rm Re}\left(A^L_{\perp}A^{R*}_{\perp}+A^L_{\parallel}A^{R*}_{\parallel}\right), \nonumber \\
I^c_1 & = & |A^L_{0}|^2+|A^R_{0}|^2 +\frac{4m^2_{\mu}}{q^2}\left[|A_t|^2 + 2 {\rm Re}\left(A^L_0 A^{R*}_0\right)\right],\nonumber \\
I_2^s &=& \frac{\beta^2_{\mu}}{4}\left[|A^L_{\perp}|^2+|A^L_{\parallel}|^2 + (L\to R)\right],\nonumber\\
I^c_2 &= & -\beta^2_{\mu} \left[|A^L_0|^2 + |A^R_0|^2\right], \nonumber \\
I_3 &= & \frac{\beta^2_{\mu}}{2} \left[|A^L_{\perp}|^2 - |A^L_{\parallel}|^2 + (L\to R)\right], \nonumber \\
I_4 &  = & \frac{\beta^2_{\mu}}{\sqrt{2}} \left[ {\rm Re}(A^L_0 A^{L*}_{\parallel}) + (L\to R)\right], \nonumber \\
I_5 &= &\sqrt{2} \beta_{\mu}\Big[{\rm Re}(A^L_0 A^{L*}_{\perp}) -(L\to R)\Big],\nonumber \\
I^s_6 &= & 2\beta_{\mu}\left[{\rm Re}(A^L_{\parallel}A^{L*}_{\perp})- (L\to R)\right],\nonumber\\
I_7 & =& \sqrt{2}\beta_{\mu}\Big[{\rm Im}(A^L_0 A^{L*}_{\parallel})- (L\to R) \Big],\nonumber\\
I_8 &= & \frac{\beta^2_{\mu}}{\sqrt{2}}\left[{\rm Im}(A^L_0A^{L*}_{\perp}) + (L\to R)\right],\nonumber \\
I_9 &= & \beta^2_{\mu}\left[{\rm Im}(A^{L*}_{\parallel}A^L_{\perp})+ (L\to R)\right].
\end{eqnarray}
\end{widetext}
The expressions of transversity amplitudes  can be found in ref.~\cite{Altmannshofer:2008dz,Bharucha:2015bzk}. These amplitudes  are written in terms of form-factors  $V(q^2)$, $A_{0,1,2}(q^2)$ and $T_{1,2,3}(q^2)$. The hadronic uncertainties in $B \to K^* \ell^+ \ell^-$ observables are mainly due to form-factors \cite{Khodjamirian:2010vf,Bharucha:2015bzk,Gubernari:2018wyi} and non-local contributions related with charm-quark loops \cite{Beneke:2001at,Khodjamirian:2010vf,Descotes-Genon:2014uoa,Capdevila:2017ert,Bobeth:2017vxj,Blake:2017fyh,Gubernari:2020eft,Gubernari:2022hxn,Gubernari:2022hxn}. The form-factors in the low-$q^2$ region are calculated using  light-cone sum rules (LCSR)  or light-meson distribution
amplitudes. In the high-$q^2$ region, the form-factors are determined from lattice computations \cite{Horgan:2013hoa,Flynn:2015ynk}.


\begin{thebibliography}{99}
\bibitem{Sakharov:1967dj}
A.~D.~Sakharov,
Pisma Zh. Eksp. Teor. Fiz. \textbf{5} (1967), 32-35

\bibitem{Kruger:1999xa}
F.~Kruger, L.~M.~Sehgal, N.~Sinha and R.~Sinha,
Phys. Rev. D \textbf{61} (2000), 114028
[erratum: Phys. Rev. D \textbf{63} (2001), 019901]
[arXiv:hep-ph/9907386 [hep-ph]].

\bibitem{Kruger:2000zg}
F.~Kruger and E.~Lunghi,
Phys. Rev. D \textbf{63} (2001), 014013
[arXiv:hep-ph/0008210 [hep-ph]].

\bibitem{Bobeth:2008ij}
C.~Bobeth, G.~Hiller and G.~Piranishvili,
JHEP \textbf{07} (2008), 106
[arXiv:0805.2525 [hep-ph]].

\bibitem{Altmannshofer:2008dz}
W.~Altmannshofer, P.~Ball, A.~Bharucha, A.~J.~Buras, D.~M.~Straub and M.~Wick,
JHEP \textbf{01} (2009), 019
[arXiv:0811.1214 [hep-ph]].

\bibitem{Bobeth:2011gi}
C.~Bobeth, G.~Hiller and D.~van Dyk,
JHEP \textbf{07} (2011), 067
[arXiv:1105.0376 [hep-ph]].



\bibitem{Alok:2011gv}
A.~K.~Alok, A.~Datta, A.~Dighe, M.~Duraisamy, D.~Ghosh and D.~London,
JHEP \textbf{11} (2011), 122
[arXiv:1103.5344 [hep-ph]].

\bibitem{Gangal:2022ole}
S.~N.~Gangal,
[arXiv:2209.02476 [hep-ph]].

\bibitem{Das:2022xjg}
D.~Das, J.~Das, G.~Kumar and N.~Sahoo,
[arXiv:2211.09065 [hep-ph]].

\bibitem{Geng:2022pld}
C.~Q.~Geng, C.~W.~Liu and Z.~Y.~Wei,
[arXiv:2212.02976 [hep-ph]].

\bibitem{Fleischer:2022klb}
R.~Fleischer, E.~Malami, A.~Rehult and K.~K.~Vos,
[arXiv:2212.09575 [hep-ph]].

  \bibitem{bsphilhc2}
R.~Aaij {\it et al.} [LHCb Collaboration],
  JHEP {\bf 1509}, 179 (2015)
  [arXiv:1506.08777 [hep-ex]].
  
    \bibitem{bsphilhc3}
R.~Aaij \textit{et al.} [LHCb],
Phys. Rev. Lett. \textbf{127} (2021) no.15, 151801
[arXiv:2105.14007 [hep-ex]].

 
\bibitem{Kstarlhcb1}
R.~Aaij {\it et al.} [LHCb Collaboration],
  Phys.\ Rev.\ Lett.\  {\bf 111}, 191801 (2013)
  [arXiv:1308.1707 [hep-ex]].
  
  
    \bibitem{Kstarlhcb2}
R.~Aaij {\it et al.} [LHCb Collaboration],
  JHEP {\bf 1602}, 104 (2016)
  [arXiv:1512.04442 [hep-ex]].
  
  \bibitem{LHCb:2020lmf}
R.~Aaij \textit{et al.} [LHCb],
Phys. Rev. Lett. \textbf{125} (2020) no.1, 011802
[arXiv:2003.04831 [hep-ex]].
  
      \bibitem{sm-angular} 
      S.~Descotes-Genon, T.~Hurth, J.~Matias and J.~Virto,
  JHEP {\bf 1305}, 137 (2013)
  [arXiv:1303.5794 [hep-ph]].
  
  \bibitem{Matias:2014jua}
J.~Matias and N.~Serra,
Phys. Rev. D \textbf{90} (2014) no.3, 034002
[arXiv:1402.6855 [hep-ph]].

\bibitem{Hofer:2015kka}
L.~Hofer and J.~Matias,
JHEP \textbf{09} (2015), 104
[arXiv:1502.00920 [hep-ph]].
  
 
\bibitem{LHCb:2021awg}
R.~Aaij \textit{et al.} [LHCb],
[arXiv:2108.09283 [hep-ex]].

\bibitem{Combination}
 [ATLAS],
ATLAS-CONF-2020-049.


\bibitem{ATLAS:2018cur}
M.~Aaboud \textit{et al.} [ATLAS],
JHEP \textbf{04} (2019), 098
[arXiv:1812.03017 [hep-ex]].


\bibitem{CMS:2019bbr}
A.~M.~Sirunyan \textit{et al.} [CMS],
JHEP \textbf{04} (2020), 188
[arXiv:1910.12127 [hep-ex]].


\bibitem{LHCb:2017rmj}
R.~Aaij \textit{et al.} [LHCb],
Phys. Rev. Lett. \textbf{118} (2017) no.19, 191801
[arXiv:1703.05747 [hep-ex]].




\bibitem{CMS:2022dbz}
 [CMS],
CMS-PAS-BPH-21-006.

\bibitem{HFLAV:2022pwe}
Y.~Amhis \textit{et al.} [HFLAV],
[arXiv:2206.07501 [hep-ex]].

\bibitem{Bobeth:2013uxa}
C.~Bobeth, M.~Gorbahn, T.~Hermann, M.~Misiak, E.~Stamou and M.~Steinhauser,
Phys. Rev. Lett. \textbf{112} (2014), 101801
[arXiv:1311.0903 [hep-ph]].

\bibitem{UTfit:2022hsi}
M.~Bona \textit{et al.} [UTfit],
[arXiv:2212.03894 [hep-ph]].


  \bibitem{Bordone:2016gaq}
M.~Bordone, G.~Isidori and A.~Pattori,
Eur. Phys. J. C \textbf{76}, no.8, 440 (2016)
[arXiv:1605.07633 [hep-ph]].

\bibitem{Hiller:2003js}
G.~Hiller and F.~Kruger,
Phys. Rev. D \textbf{69} (2004), 074020
[arXiv:hep-ph/0310219 [hep-ph]].

\bibitem{LHCb:2021trn}
R.~Aaij \textit{et al.} [LHCb],
Nature Phys. \textbf{18}, no.3, 277-282 (2022)
[arXiv:2103.11769 [hep-ex]].

\bibitem{rkstar}
R.~Aaij \textit{et al.} [LHCb],
JHEP \textbf{08} (2017), 055
[arXiv:1705.05802 [hep-ex]].

\bibitem{Isidori:2020acz}
G.~Isidori, S.~Nabeebaccus and R.~Zwicky,
JHEP \textbf{12}, 104 (2020)
[arXiv:2009.00929 [hep-ph]].

\bibitem{Isidori:2022bzw}
G.~Isidori, D.~Lancierini, S.~Nabeebaccus and R.~Zwicky,
JHEP \textbf{10}, 146 (2022)
[arXiv:2205.08635 [hep-ph]].

\bibitem{Nabeebaccus:2022pje}
S.~Nabeebaccus and R.~Zwicky,
[arXiv:2209.09585 [hep-ph]].

\bibitem{Ciuchini:2015qxb}
M.~Ciuchini, M.~Fedele, E.~Franco, S.~Mishima, A.~Paul, L.~Silvestrini and M.~Valli,
JHEP \textbf{06} (2016), 116
[arXiv:1512.07157 [hep-ph]].

\bibitem{Hurth:2016fbr}
T.~Hurth, F.~Mahmoudi and S.~Neshatpour,
Nucl. Phys. B \textbf{909} (2016), 737-777
[arXiv:1603.00865 [hep-ph]].

\bibitem{Ciuchini:2021smi}
M.~Ciuchini, M.~Fedele, E.~Franco, A.~Paul, L.~Silvestrini and M.~Valli,
Eur. Phys. J. C \textbf{83} (2023) no.1, 64
[arXiv:2110.10126 [hep-ph]].


\bibitem{Ciuchini:2022wbq}
M.~Ciuchini, M.~Fedele, E.~Franco, A.~Paul, L.~Silvestrini and M.~Valli,
[arXiv:2212.10516 [hep-ph]].

\bibitem{LHCb:2022qnv}
 [LHCb],
[arXiv:2212.09152 [hep-ex]].
  
      \bibitem{LHCb:2022zom}
 [LHCb],
[arXiv:2212.09153 [hep-ex]].

\bibitem{LHCb:2021lvy}
R.~Aaij \textit{et al.} [LHCb],
[arXiv:2110.09501 [hep-ex]].

  
  
      
  \bibitem{Descotes-Genon:2013wba}
S.~Descotes-Genon, J.~Matias and J.~Virto,
Phys. Rev. D \textbf{88}, 074002 (2013)
[arXiv:1307.5683 [hep-ph]].

\bibitem{Altmannshofer:2013foa}
W.~Altmannshofer and D.~M.~Straub,
Eur. Phys. J. C \textbf{73}, 2646 (2013)
[arXiv:1308.1501 [hep-ph]].

\bibitem{Hurth:2013ssa}
T.~Hurth and F.~Mahmoudi,
JHEP \textbf{04}, 097 (2014)
[arXiv:1312.5267 [hep-ph]].

\bibitem{Capdevila:2016ivx}
B.~Capdevila, S.~Descotes-Genon, J.~Matias and J.~Virto,
JHEP \textbf{10}, 075 (2016)
[arXiv:1605.03156 [hep-ph]].

\bibitem{Ciuchini:2017mik}
M.~Ciuchini, A.~M.~Coutinho, M.~Fedele, E.~Franco, A.~Paul, L.~Silvestrini and M.~Valli,
Eur. Phys. J. C \textbf{77} (2017) no.10, 688
[arXiv:1704.05447 [hep-ph]].

\bibitem{Alok:2017jgr}
A.~K.~Alok, B.~Bhattacharya, D.~Kumar, J.~Kumar, D.~London and S.~U.~Sankar,
Phys. Rev. D \textbf{96} (2017) no.1, 015034
[arXiv:1703.09247 [hep-ph]].

 
 \bibitem{Alok:2019ufo}
A.~K.~Alok, A.~Dighe, S.~Gangal and D.~Kumar,
JHEP \textbf{06} (2019), 089
[arXiv:1903.09617 [hep-ph]].


\bibitem{Altmannshofer:2021qrr}
W.~Altmannshofer and P.~Stangl,
Eur. Phys. J. C \textbf{81} (2021) no.10, 952
[arXiv:2103.13370 [hep-ph]].

\bibitem{Carvunis:2021jga}
A.~Carvunis, F.~Dettori, S.~Gangal, D.~Guadagnoli and C.~Normand,
[arXiv:2102.13390 [hep-ph]]

\bibitem{Alguero:2021anc}
M.~Alguer\'o, B.~Capdevila, S.~Descotes-Genon, J.~Matias and M.~Novoa-Brunet,
[arXiv:2104.08921 [hep-ph]]

\bibitem{Geng:2021nhg}
L.~S.~Geng, B.~Grinstein, S.~J\"ager, S.~Y.~Li, J.~Martin Camalich and R.~X.~Shi,
[arXiv:2103.12738 [hep-ph]]

\bibitem{Hurth:2021nsi}
T.~Hurth, F.~Mahmoudi, D.~M.~Santos and S.~Neshatpour,
[arXiv:2104.10058 [hep-ph]].


\bibitem{Angelescu:2021lln}
A.~Angelescu, D.~Be\v{c}irevi\'c, D.~A.~Faroughy, F.~Jaffredo and O.~Sumensari,
Phys. Rev. D \textbf{104}, no.5, 055017 (2021)
[arXiv:2103.12504 [hep-ph]].

\bibitem{Alok:2022pjb}
A.~K.~Alok, N.~R.~Singh Chundawat, S.~Gangal and D.~Kumar,
Eur. Phys. J. C \textbf{82} (2022) no.10, 967
[arXiv:2203.13217 [hep-ph]].

\bibitem{SinghChundawat:2022ldm}
N.~R.~Singh Chundawat,
To appear in Phys. Rev. D (2023)
[arXiv:2212.01229 [hep-ph]].

  \bibitem{LHCb:2021xxq}
R.~Aaij \textit{et al.} [LHCb],
JHEP \textbf{11}, 043 (2021)
[arXiv:2107.13428 [hep-ex]]

\bibitem{LHCb:2014vgu}
R.~Aaij \textit{et al.} [LHCb],
Phys. Rev. Lett. \textbf{113}, 151601 (2014)
[arXiv:1406.6482 [hep-ex]].



  \bibitem{Straub:2018kue} 
  D.~M.~Straub,
  arXiv:1810.08132 [hep-ph].
  


\bibitem{Belle:2019oag}
A.~Abdesselam \textit{et al.} [Belle],
Phys. Rev. Lett. \textbf{126}, no.16, 161801 (2021)
[arXiv:1904.02440 [hep-ex]].


\bibitem{Lees:2013nxa}
J.~P.~Lees {\it et al.} [BaBar Collaboration],
  Phys.\ Rev.\ Lett.\  {\bf 112}, 211802 (2014)
  [arXiv:1312.5364 [hep-ex]].


\bibitem{LHCb:2016ykl}
R.~Aaij \textit{et al.} [LHCb],
JHEP \textbf{11} (2016), 047
[arXiv:1606.04731 [hep-ex]].



\bibitem{Khachatryan:2015isa}
  V.~Khachatryan {\it et al.} [CMS Collaboration],
  Phys.\ Lett.\ B {\bf 753}, 424 (2016)
  [arXiv:1507.08126 [hep-ex]].

  \bibitem{CDFupdate}
CDF Collaboration, 
 CDF public note 10894.


  \bibitem{Aaij:2014pli}
R.~Aaij {\it et al.} [LHCb Collaboration],
  JHEP {\bf 1406}, 133 (2014)
  [arXiv:1403.8044 [hep-ex]].

  
  
  \bibitem{ATLAS:2018gqc}
M.~Aaboud \textit{et al.} [ATLAS],
JHEP \textbf{10}, 047 (2018)
[arXiv:1805.04000 [hep-ex]].
  
  \bibitem{CMS:2017rzx}
A.~M.~Sirunyan \textit{et al.} [CMS],
Phys. Lett. B \textbf{781}, 517-541 (2018)
[arXiv:1710.02846 [hep-ex]].

  \bibitem{LHCb:2020gog}
R.~Aaij \textit{et al.} [LHCb],
Phys. Rev. Lett. \textbf{126}, no.16, 161802 (2021)
[arXiv:2012.13241 [hep-ex]].
 

\bibitem{LHCb:2015ycz}
R.~Aaij \textit{et al.} [LHCb],
JHEP \textbf{04}, 064 (2015)
[arXiv:1501.03038 [hep-ex]].

\bibitem{Belle:2016fev}
S.~Wehle \textit{et al.} [Belle],
Phys. Rev. Lett. \textbf{118} (2017) no.11, 111801
[arXiv:1612.05014 [hep-ex]].


\bibitem{James:1975dr}
F.~James and M.~Roos,
Comput. Phys. Commun. \textbf{10}, 343-367 (1975)

\bibitem{Bharucha:2015bzk}
A.~Bharucha, D.~M.~Straub and R.~Zwicky,
JHEP \textbf{08}, 098 (2016)
[arXiv:1503.05534 [hep-ph]].

\bibitem{Gubernari:2018wyi}
N.~Gubernari, A.~Kokulu and D.~van Dyk,
JHEP \textbf{01}, 150 (2019)
d
[arXiv:1811.00983 [hep-ph]].




\bibitem{dattlon}
A.~Datta and D.~London,
  Phys.\ Lett.\ B {\bf 595}, 453 (2004)
  [hep-ph/0404130].


\bibitem{Belle:2009zue}
J.~T.~Wei \textit{et al.} [Belle],
Phys. Rev. Lett. \textbf{103}, 171801 (2009)
[arXiv:0904.0770 [hep-ex]].

\bibitem{BaBar:2012mrf}
J.~P.~Lees \textit{et al.} [BaBar],
Phys. Rev. D \textbf{86}, 032012 (2012)
[arXiv:1204.3933 [hep-ex]].

\bibitem{LHCb:2014mit}
R.~Aaij \textit{et al.} [LHCb],
JHEP \textbf{09}, 177 (2014)
[arXiv:1408.0978 [hep-ex]].

\bibitem{LHCb:2012kz}
R.~Aaij \textit{et al.} [LHCb],
Phys. Rev. Lett. \textbf{110}, no.3, 031801 (2013)
[arXiv:1210.4492 [hep-ex]].

\bibitem{LHCb:2013lvw}
R.~Aaij \textit{et al.} [LHCb],
Phys. Rev. Lett. \textbf{111}, no.15, 151801 (2013)
[arXiv:1308.1340 [hep-ex]].

\bibitem{Belle-II:2018jsg}
E.~Kou \textit{et al.} [Belle-II],
PTEP \textbf{2019}, no.12, 123C01 (2019)
[erratum: PTEP \textbf{2020}, no.2, 029201 (2020)]
[arXiv:1808.10567 [hep-ex]].
  
  \bibitem{Gratrex:2015hna}
J.~Gratrex, M.~Hopfer and R.~Zwicky,
Phys. Rev. D \textbf{93}, no.5, 054008 (2016)
[arXiv:1506.03970 [hep-ph]].

\bibitem{Cerri:2018ypt}
A.~Cerri, V.~V.~Gligorov, S.~Malvezzi, J.~Martin Camalich, J.~Zupan, S.~Akar, J.~Alimena, B.~C.~Allanach, W.~Altmannshofer and L.~Anderlini, \textit{et al.}
CERN Yellow Rep. Monogr. \textbf{7}, 867-1158 (2019)
[arXiv:1812.07638 [hep-ph]].


\bibitem{Becirevic:2012fy}
D.~Becirevic, N.~Kosnik, F.~Mescia and E.~Schneider,
Phys. Rev. D \textbf{86}, 034034 (2012)
[arXiv:1205.5811 [hep-ph]].

\bibitem{Bobeth:2007dw}
C.~Bobeth, G.~Hiller and G.~Piranishvili,
JHEP \textbf{12}, 040 (2007)
[arXiv:0709.4174 [hep-ph]].



\bibitem{Charles:1998dr}
J.~Charles, A.~Le Yaouanc, L.~Oliver, O.~Pene and J.~C.~Raynal,
Phys. Rev. D \textbf{60}, 014001 (1999)
[arXiv:hep-ph/9812358 [hep-ph]].

\bibitem{Beneke:2000wa}
M.~Beneke and T.~Feldmann,
Nucl. Phys. B \textbf{592}, 3-34 (2001)
[arXiv:hep-ph/0008255 [hep-ph]].

\bibitem{Bobeth:2011nj}
C.~Bobeth, G.~Hiller, D.~van Dyk and C.~Wacker,
JHEP \textbf{01}, 107 (2012)
[arXiv:1111.2558 [hep-ph]].



\bibitem{Khodjamirian:2010vf}
A.~Khodjamirian, T.~Mannel, A.~A.~Pivovarov and Y.~M.~Wang,
JHEP \textbf{09}, 089 (2010)
[arXiv:1006.4945 [hep-ph]].



\bibitem{Beneke:2001at}
M.~Beneke, T.~Feldmann and D.~Seidel,
Nucl. Phys. B \textbf{612}, 25-58 (2001)
[arXiv:hep-ph/0106067 [hep-ph]].


\bibitem{Descotes-Genon:2014uoa}
S.~Descotes-Genon, L.~Hofer, J.~Matias and J.~Virto,
JHEP \textbf{12}, 125 (2014)
[arXiv:1407.8526 [hep-ph]].

\bibitem{Capdevila:2017ert}
B.~Capdevila, S.~Descotes-Genon, L.~Hofer and J.~Matias,
JHEP \textbf{04}, 016 (2017)
[arXiv:1701.08672 [hep-ph]].

\bibitem{Bobeth:2017vxj}
C.~Bobeth, M.~Chrzaszcz, D.~van Dyk and J.~Virto,
Eur. Phys. J. C \textbf{78}, no.6, 451 (2018)
[arXiv:1707.07305 [hep-ph]].

\bibitem{Blake:2017fyh}
T.~Blake, U.~Egede, P.~Owen, K.~A.~Petridis and G.~Pomery,
Eur. Phys. J. C \textbf{78}, no.6, 453 (2018)
[arXiv:1709.03921 [hep-ph]].

\bibitem{Gubernari:2020eft}
N.~Gubernari, D.~van Dyk and J.~Virto,
JHEP \textbf{02}, 088 (2021)
[arXiv:2011.09813 [hep-ph]].

\bibitem{Gubernari:2022hxn}
N.~Gubernari, M.~Reboud, D.~van Dyk and J.~Virto,
JHEP \textbf{09} (2022), 133
[arXiv:2206.03797 [hep-ph]].

\bibitem{Horgan:2013hoa}
R.~R.~Horgan, Z.~Liu, S.~Meinel and M.~Wingate,
Phys. Rev. D \textbf{89}, no.9, 094501 (2014)
[arXiv:1310.3722 [hep-lat]].

\bibitem{Flynn:2015ynk}
J.~Flynn, A.~J\"uttner, T.~Kawanai, E.~Lizarazo and O.~Witzel,
PoS \textbf{LATTICE2015}, 345 (2016)
[arXiv:1511.06622 [hep-lat]].

\end{thebibliography}
\end{document}